\documentclass[aps,prd,twocolumn,superscriptaddress,showpacs,a4paper]{revtex4}
\usepackage{graphicx}
\usepackage{dcolumn}
\usepackage{bm}
\usepackage{natbib}

\newcommand{\be}{\begin{equation}}
\newcommand{\ee}{\end{equation}}
\newcommand{\bea}{\begin{eqnarray}}
\newcommand{\eea}{\end{eqnarray}}

\newcommand{\WMAP}{{\slshape WMAP~}}
\newcommand{\LCDM}{$ \Lambda $CDM~}

\begin{document}

\title{Combined analysis of the integrated Sachs--Wolfe effect and cosmological implications}

\author {Tommaso~Giannantonio}

\affiliation {Institute of Cosmology and Gravitation, University
of Portsmouth, Mercantile House, Hampshire Terrace, Portsmouth PO1 2EG, UK}

\author {Ryan~Scranton} 
\affiliation {Department of Physics and Astronomy, University of Pittsburgh, 3941 O'Hara St., Pittsburgh, PA 15260, USA}

\author {Robert~G.~Crittenden}

\affiliation {Institute of Cosmology and Gravitation, University
of Portsmouth, Mercantile House, Hampshire Terrace, Portsmouth PO1 2EG, UK}

\author {Robert~C.~Nichol}

\affiliation {Institute of Cosmology and Gravitation, University
of Portsmouth, Mercantile House, Hampshire Terrace, Portsmouth PO1 2EG, UK}

\author {Stephen~P.~Boughn}
\affiliation {Haverford College, Haverford PA 19041, USA}

\author {Adam~D.~Myers} 
\affiliation {Department of Astronomy, University of Illinois at Urbana--Champaign, Urbana, IL 61820, USA}

\author {Gordon~T.~Richards} 
\affiliation {Department of Physics, Drexel University, 3141 Chestnut Street, Philadelphia, PA 19104, USA}

\begin {abstract}
We present a global measurement of the integrated Sachs--Wolfe (ISW) effect
obtained by cross-correlating all relevant large scale galaxy data sets
with the cosmic microwave background radiation map provided by the Wilkinson
Microwave Anisotropy Probe.  With these measurements, the overall ISW signal is detected at the $\sim 4.5 \sigma$ level.  We also examine the 
cosmological implications of these measurements, particularly the dark energy equation of state $w$, its 
sound speed $c_s$, and the overall
curvature of the Universe. 
The flat \LCDM model is a good fit to  
the data and, assuming this model, we find that the ISW data  
constrain $\Omega_m = 0.20^{+0.19}_{-0.11} $ at the $ 95\% $ confidence level.
When we combine our ISW results with the latest baryon oscillation and supernovae  
measurements, we find that the result is still consistent with a flat \LCDM model with $ w=-1$ out to redshifts $ z > 1$.

\end {abstract}

\pacs {98.80.Es, 98.54.Aj, 98.70.Vc}

\maketitle

\section {Introduction} \label {sec:intro}
There is growing evidence that the expansion of the Universe
is dominated by an unknown \emph{dark energy}, which 
accounts for more than 70\% of the total matter that we observe. 
Measurements of the Hubble diagram of type Ia supernovae
\cite{Riess:2004nr, Astier:2005qq} indicate this dark energy 
is causing the Universe's expansion to accelerate, while cosmic microwave
background (CMB) anisotropies power spectrum experiments, such as \WMAP
\cite{Hinshaw:2006ia}, limit the amount of spatial curvature which might 
otherwise explain the low dark matter density.  Understanding the nature of 
this dark energy is essential, and many possible explanations have been proposed, 
from a cosmological constant, to quintessence, to modifying the laws of gravity. 

To discriminate between possible explanations of dark energy, many alternative 
probes have been developed, including measuring the Universe's geometry through  
observations of the baryon oscillation (BAO) scale \cite{Eisenstein:2005su,Cole:2005sx,Percival:2007yw} and probing the 
dark matter density through observations of clusters of galaxies \cite{Allen:2004cd,Allen:2007ue}. 
Here we consider another impact of dark energy, the creation of CMB anisotropies 
at late times via the  
integrated Sachs--Wolfe (ISW) effect~\cite{Sachs:1967er}.  While most CMB anisotropies 
are generated at very early times, further fluctuations can be induced gravitationally 
at late times as photons pass through evolving gravitational potentials.  If dark matter 
dominates, the gravitational potentials do not vary with time, but the presence of 
dark energy or spatial curvature will cause the potentials to evolve at late times, producing new temperature 
fluctuations at low redshifts (primarily at $z < 2$).   

Directly observing these new CMB temperature anisotropies is challenging, primarily 
because their amplitudes are a fraction of the anisotropies arising from higher redshifts. 
They are also predominantly seen on large scales, where the uncertainty in the observations 
from cosmic variance is biggest.  The search for the ISW effect has instead focused on finding correlations 
between the CMB temperature maps and maps of the density which trace the local gravitational 
potentials \cite{Crittenden:1995ak}.   Primordial anisotropies should be uncorrelated 
with the local density, making it possible to pull out the weaker ISW anisotropies.  

Many groups have detected this correlation using the accurate \WMAP CMB
data and various density probes distributed at a range of redshifts and in
different regions of the electromagnetic spectrum: from shallow infrared 
observations~\cite{Afshordi:2003xu, Rassat:2006kq}, to optical surveys such
as the APM and the SDSS~\cite{Scranton:2003in, Fosalba:2003iy,
Fosalba:2003ge, Padmanabhan:2004fy, Cabre:2006qm}, radio galaxy catalogues 
\cite{Boughn:2003yz, Nolta:2003uy}, X-ray surveys~\cite{Boughn:2003yz},
and the deepest quasars from the SDSS~\cite{Giannantonio:2006du}. 
These measures span a range of redshift going from $ z = 0.1 $ to 
$ z = 1.5 $, where the ISW effect has been measured at significance levels 
typically around $ 2-3 \sigma $, and appear generally compatible with the
expectation from the \LCDM model.

Although indicative of the presence of dark energy, none of these measures
alone has significant power to constrain models due to their low
significance.  Thus, it is important to combine the various observations; but some care must 
be taken in doing so.  The surveys are often overlapping both in sky coverage and in redshift 
range, meaning there are likely covariances between them that may be important 
when considering a large scale effect like the ISW.  In addition, 
these measurements have been made with a variety of techniques, using 
angular correlations, Fourier modes, or a range of wavelet techniques \cite{Vielva:2004zg,McEwen:2006my,Pietrobon:2006gh,McEwen:2007rz}.
The error bars themselves have also been estimated using different techniques, using both jack-knife 
approaches and Monte Carlo simulations of the CMB sky.    

A combined analysis has been attempted in the
past adding several measures in order to extend the constraining power
in redshift and learn more about the behaviour of dark energy and other
cosmological parameters~\cite{Gaztanaga:2004sk, Corasaniti:2005pq,
Cooray:2005px, Giannantonio:2006ij}.  However, this analysis largely ignored 
the differences in the observations and accounted for the covariances between experiments in 
a fairly arbitrary way.   Here we perform a combined analysis by reanalysing all the observations 
in a consistent way, measuring directly the covariances between the different observations using 
a number of different methods and looking at the cross-correlations between all the various data sets.  In this way
we hope to give a definitive result for the ISW evidence for dark energy and the resulting cosmological 
constraints.  

This paper is structured as follows: we begin in section \S\ref{sec:theo} by giving a brief theoretical 
description of the ISW effect, how the cross-correlation is measured, and the 
important issue of estimating the covariance between observations.  
In section \S\ref{sec:cats} we describe
the catalogues used for the cross-correlation, and in section \S\ref{sec:results} we show the measurements of the various cross-correlation functions between the different catalogues and their
cross-correlation with the CMB.   We discuss the significance of the measurements in section \S\ref{sec:signif} and show the resulting cosmological constraints in section  \S\ref{sec:constraints}, before some concluding remarks in section \S\ref{sec:conclusion}.

\section {Method}  \label {sec:theo}

\subsection {The ISW effect}

We begin with some theoretical background on the ISW effect and its 
detection.   Most of the CMB anisotropies we observe were created at 
the redshift of last scattering ($ z \sim 1100 $) when the Universe was 400 ky old, 
as a result of fluctuations in the photon density, velocity and potential energy.  
Since that time, the CMB photons have travelled largely untouched, but anisotropies 
can still be produced gravitationally if the photons pass through time varying potential 
wells, 
\be
\frac {\Delta T} {T} (\hat \mathbf{n}) = - 2 \int \dot \Phi \left[ \tau, \hat \mathbf{n} (\tau_0 - \tau)  \right] d \tau,
\ee
where $ \tau $ is the conformal time, the dot represents a conformal
time derivative and the integral is intended along the line of sight of
the photon.  (Throughout, we work in units where the speed of light is unity and for 
simplicity assume that the effects of anisotropic stress can be ignored).   When the gravitational 
potential decays due to its linear evolution, this is usually referred to as the integrated Sachs-Wolfe effect, and 
if the potential decay is a result of non-linear evolution, as in clusters, it is referred to as the Rees-Sciama effect \cite{Rees:1968}.    

The physical picture is very straightforward; as a CMB photon falls into a gravitational potential well, it gains energy; as the photon climbs out of a potential well, it loses energy.
These effects exactly cancel if the potential is time independent, but can result in a net kick if the potential 
evolves as the photon passes through it.  In particular, we know that during the matter dominated era the gravitational potential 
stays constant, and so $ \dot \Phi = 0 $, which means that in that era there will not be any ISW produced. This
changes if dark energy or curvature become important at later times: in this case, $\dot \Phi \neq 0$ and additional CMB anisotropies will be produced.
In the usual case, the potential amplitudes decrease at late times, so that a temperature increase results from passing through potential wells, while a temperature deficits results from traversing potential hills.  

Since we know from the CMB experiments such as \WMAP that the Universe is
very close to flat, we can attribute most of the late ISW to dark energy;
observing it can therefore provide a probe of dark energy, its properties
and its evolution in time.  In particular, the ISW effect could be one of the few 
ways of measuring the sound speed of dark energy \cite{Hu:2004yd}.  However, it can also provide useful information about the curvature of the Universe \cite{Kinkhabwala:1998zj}.

Unfortunately, measurement of the ISW effect is made difficult by the
embedding of the small ISW signal in the much larger (10 times) primary CMB
anisotropies. Furthermore, the total ISW signal is due to all the density fluctuations, both positive and negative, 
along the line of sight.   On small scales, the individual temperature differences are small and they tend to cancel out.  
The most significant ISW effect results from the coherent large scale potentials, but unfortunately these scales are precisely where 
cosmic variance is most troublesome. 

This problem can be overcome by examining how the ISW temperature correlates with the 
density of galaxies, which should trace the potential wells and hills which bring about the anisotropies.  
The observed galaxy density contrast in a given direction $ \hat \mathbf {n}_1 $ 
will be
\be
\delta_g (\hat \mathbf {n}_1) =  \int b_g (z) \frac{dN}{dz}(z) \delta_m (\hat \mathbf {n}_1, z) dz,
\ee
where $ {dN}/{dz} $ is the selection function of the survey, $ b_g (z)$ its 
galaxy bias relating the visible matter distribution to the underlying dark 
matter, and $ \delta_m $ the matter density perturbations.      Since the density $\delta_m$ is related to the potential $\Phi$ by the Poisson equation, the observed galaxy density will be correlated with the ISW temperature fluctuation in the nearby direction $ \hat \mathbf {n}_2 $, which is 
\be
\frac{\Delta T}{T} (\hat \mathbf {n}_2) = -2 \int e^{-\tau(z)} \frac {d\Phi} {dz} (\hat \mathbf {n}_2, z) dz,
\ee
where $ e^{-\tau(z)} $ is the visibility function of the photons, which accounts for the effect of photons re-scattering following reionisation.

The galaxy bias, $b_g(z)$, can evolve in time or as a function of scale; however, we will generally assume that it is time and scale independent for simplicity.  For our purposes, a time dependent bias is equivalent to changing the selection function of the survey.  Scale dependence of the bias is more problematic, but on the very large scales ($> 10$ Mpc) we are considering, the scale dependence is expected to be weak \cite{Blanton:1998aa,Percival:2006gt}. 

Given a map of the CMB and a survey of galaxies, the angular 
auto-correlation and cross-correlation functions are defined as
\bea
C^{Tg}(\vartheta) & \equiv & \langle \frac{\Delta T}{T} (\hat \mathbf {n}_1) \delta_g (\hat \mathbf {n}_2)  \rangle \\
C^{gg}(\vartheta) & \equiv & \langle \delta_g (\hat \mathbf {n}_1) \delta_g (\hat \mathbf {n}_2)  \rangle,
\eea
with the average carried over all the pairs at the same angular 
distance $ \vartheta = | \hat \mathbf {n}_1 - \hat \mathbf {n}_2 |$.

It is possible to express these quantities in the harmonic space with the 
use of the Legendre polynomials $P_l$:
\be
C^{Tg}(\vartheta) = \sum_{l=2}^{\infty} \frac {2 l + 1} {4 \pi} C_l^{Tg} 
P_l [\cos (\vartheta)],
\ee
and the auto- and cross-correlation power spectra are given by
\bea
C_l^{Tg} & = & 4 \pi \int \frac {dk}{k} \Delta^2(k) I_l^{{ISW}} (k) 
I_l^{g} (k) \\
C_l^{gg} & = & 4 \pi \int \frac {dk}{k} \Delta^2(k) I_l^g (k) I_l^{g} (k),
\eea
where $ \Delta(k) $ is the scale invariant matter power spectrum 
$ \Delta^2 (k) \equiv 4 \pi k^3 P(k) / (2 \pi)^3 $ and the two integrands 
are respectively
\bea
I_l^{ISW}(k) & = & -2 \int e^{-\tau(z)} \frac {d\Phi_k} {dz} 
j_l[k \chi(z)] dz \\
I_l^{g}(k)   & = & \int b_g(z) \frac{dN}{dz} (z) \delta_m (k,z) j_l[k \chi(z)] dz,
\eea
where $ \Phi_k, \delta_m(k,z) $ are the Fourier components of the gravitational 
potential and matter perturbations, $ j_l (x) $ are the spherical Bessel functions 
and $ \chi $ is the comoving distance.  These are calculated using a modified version of {\tt cmbfast} \cite{Seljak:1996is,Corasaniti:2005pq}. 

\subsection {Theoretical signal-to-noise}

Unfortunately, the ability to detect the cross-correlation is limited because the signal falls off on small 
scales.  Not only is cosmic variance an important factor, but there is also the problem of accidental 
correlations between the galaxy surveys and the CMB anisotropies produced at last scattering.  Many independent measurements 
are needed to reduce the impact of such accidental correlations.  
The signal to noise ratio of the 
CCF with a particular survey is given by
\be
\left( \frac{S}{N} \right)^2 = \sum_l (2l+1) \frac { [C^{Tg}_l]^2} {C_l^{gg} C_l^{TT} + [C_l^{Tg}]^2}.
\ee
For the ISW, we are usually in the weak correlation regime, so that $ C_l^{Tg} /\sqrt{C_l^{gg} C_l^{TT}} \ll 1 .$

The signal to noise can be separated to obtain the contribution as a function of redshift; for a standard \LCDM cosmology, most of the signal is expected at $ z < 3 $, with the peak around a redshift of $z \simeq 0.5$ \cite{Afshordi:2004kz}.  While the signal is highest at low redshifts, more independent volumes are available for higher $z$.  The signal  to noise scales roughly as the square root of the fraction of the sky observed.  

The most optimistic case is when the distribution of galaxies follows precisely the evolution of the 
ISW effect.  In this case, $C_l^{Tg} =  C_l^{gg} = C_l^{ISW}$ where $C_l^{ISW}$ is the spectrum of the ISW temperature anisotropies alone, which is assumed to 
be much smaller than the total CMB anisotropy, $C_l^{ISW} \ll C_l^{TT}$.
Thus, the signal to noise reduces to~\cite {Crittenden:1995ak} 
\be
\left( \frac{S}{N} \right)^2 \simeq \sum_{l}  (2 l +1) \frac{C_l^{ISW}}{C_l^{TT}},
\ee 
This gives an optimistic total $ S/N \simeq 7-10 $ for a standard \LCDM cosmology.
The ISW constraints which might arise from realistic future surveys can be found in \cite{Pogosian:2005ez}.

\subsection {Correlation estimators}

Our aim is to estimate the correlation between several galaxy surveys and 
the CMB: as described above, this measure can be performed in the real space 
using the CCF or in the harmonic space with the cross-correlation power 
spectrum. The two methods are theoretically equivalent for a full sky 
analysis and both have been used to detect the ISW cross-correlations. However, when one moves 
away from the ideal full-sky scenario, it is more straightforward to account for the sky mask using the real space correlations, and therefore we will follow this approach here.

The matter density and CMB temperature as well 
as their projections onto the celestial sphere are in principle continuous fields; however, 
we only have access to the sampling of these fields experimentally obtained 
by measuring the CMB temperature in some fixed directions and counting the 
number of galaxies in a given patch of sky.   
In practise, we pixelise these maps using the HEALPix pixelisation scheme~\cite {Gorski:2004by}, using a relatively coarse 
resolution: $ N_\mathrm{side} = 64 $, corresponding to $ N_\mathrm{pix} = $ 49,152 pixels with dimensions 
$ 0.92^{\circ} \times 0.92^{\circ} $.  This resolution is sufficient for a large scale correlations like the ISW and makes 
it tractable to perform large numbers of Monte Carlo simulations. A finer resolution ($ N_\mathrm{side} = 128 $) was explored, but the results did not change significantly.

In making the maps, we assign the average temperature or the total number of galaxies to each pixel.  The maps are masked according to the 
particular requirements for each catalogue and the most relevant foregrounds as discussed below.
It is inevitable that some pixels are only partially covered in the original survey, either because only part of the area was observed, 
or because some of this area was masked out.  In such cases, predominately occurring on the edge of the survey, the number of galaxies in a pixel is estimated as $n_i = n^{\mathrm{obs}}_i /f_i$ where $f_i$ is the fraction of the pixel observed.   The mask was obtained through 
sampling all objects in each catalogue in a higher resolution ($ N_{\mathrm{high}} = 512 $)
as described in~\cite{Giannantonio:2006du}.

From these maps, both the auto- and cross-correlations were estimated, down-weighting those pixels with 
partial coverage proportionally to $f_i$.  For the auto-correlation functions (ACFs), we used the estimator, 
\be
\hat{C} (\vartheta) = \frac {1} {N_{\vartheta}} \sum_{i,j}  f_i \left( n_i - \bar n \right) f_j \left( n_j - \bar n \right),
\ee
where $ \bar n$ is the average number of galaxies in a pixel for the survey of interest,  and
$ N_{\vartheta} = \sum_{ij} f_i f_j $ is the weighted number of pairs of pixels with separation $ \vartheta $.  For the temperature maps, 
we simply replace $n_i$ and $\bar{n}$ with the pixel temperature and average temperature of the CMB maps.   

More generally, we are interested in the cross-correlation function between the survey $p$ and the survey $q$; this is estimated 
similarly, accounting for the fact that the pixel weighting and mean number per pixel will depend on the survey,  
\be
\hat{C}^{pq} (\vartheta) = \frac {1} {N_{\vartheta}^{pq}} \sum_{i,j}  f_i^p \left( n_i^p - \bar n^p \right) f_j^q \left( n_j^q - \bar n^q \right).
\ee
The number of pairs of pixels at a given separation,
$ N_{\vartheta}^{pq} = \sum_{ij} f_i^p f_j^q ,$ will depend on both of the surveys under consideration.  
This again extends to the density-CMB CCFs in the obvious way.  

We use $ N_b = 13 $ 
angular bins in the range $ 0^{\circ} < \vartheta < 12^{\circ} $.  We use a linear binning, and have explored 
the dependence of our results on the choice of binning, changing both the number and trying a logarithmic binning; 
neither had significant impact on the results.  

\subsection{Covariance estimators} 

An important aspect of this calculation is the estimation of the covariance of the 
cross-correlation measurements. As described most recently by~\cite{Cabre:2007rv}, there are a number of different ways 
to calculate the errors on this measurement, each with their own advantages and drawbacks.  Here, we calculate 
our errors in three ways: a Monte Carlo method (MC1), where the covariance matrix is estimated by measuring the CCF between random CMB 
maps while keeping fixed the observed density map; a second Monte Carlo method (MC2), similar to MC1  
but including also random density maps which are correlated at the expected level with the random temperature maps; and jack-knife errors (JK) which are estimated by looking at the variance of the CCF when patches of the sky are removed. 

The first approach is to generate random Monte Carlo maps of the CMB assuming 
the \WMAP best fit cosmology, and estimating the covariance matrix 
cross-correlating these maps with the true density maps (MC1).  

The \WMAP third year fiducial model we use throughout this paper has baryon density $\Omega_b = 0.04185 $, matter density $ \Omega_m = 0.2402 $, Hubble constant $ H_0 = 73.0 $, scalar spectral index $ n_s = 0.958 $, optical depth $ \tau = 0.092 $ and amplitude of density fluctuations $ A = 0.80 $ at $ k = 0.002 $ Mpc$^{-1}$.

  The MC1 is the most widely used estimator in the 
literature, though here we extend the usual calculation to account for covariances between the CCFs of 
the CMB with different surveys.  
This method is reasonably fast to implement and accounts for the cosmic variance 
and the accidental correlations with the CMB which are the primary source of error.   
However, it is asymmetrical, in that it does not account for the variance in 
the density maps or its Poisson noise; 
the MC1 method also assumes there are no cross-correlations, 
though the expected (and observed) weakness 
of the cross-correlation indicate that this should not introduce a large bias. 
Finally, like all Monte Carlo approaches, it is model dependent and could fail if the data model is 
poorly understood (e.g. foregrounds or non-Gaussianity of the maps). 

However, some of these problems can be avoided if we also generate 
random density maps for each catalogue based on the \WMAP cosmology and the 
redshift distribution, with the addition of Poisson noise to the maps (MC2). In this 
case, we have the ability to account for the expected correlations between the maps as 
described in Appendix \ref {app:maps}. This method is more time demanding, in that it requires more 
random maps for each correlation measurement; it also retains the unwanted 
model dependence, and unlike the previous method has no explicit dependence on any of the observed maps.  

To estimate the covariance between the different angular bins of a single CCF 
following the MC1 and MC2 methods for each catalogue $k$ we use the following 
estimator of the full covariance matrix:
\be \label {eq:covestMC}
{\cal {C}}_{ij} = \frac {1} {M} \sum_{k = 1}^{M} \left[ \hat C^{Tg}_{k} (\vartheta_i) - \bar C^{Tg} (\vartheta_i) \right] \left[ \hat C^{Tg}_{k} (\vartheta_j) - \bar C^{Tg} (\vartheta_j) \right],
\ee
where $ \bar C^{Tg} (\vartheta_i) $ are the mean correlation 
functions in the $i$-th angular bin over $M$ realisations; the diagonal part 
of these matrices gives the variance of the CCF in each bin,
${\cal {C}}_{ii}^k = \sigma^2_i,$
while the off-diagonal part represents the covariance between the points.

The last method to estimate the covariance (jack-knife) consists in estimating the 
variance by generating mock density maps from the true ones, simply discarding 
a small patch of them. In practise, we can divide the original density map 
in $M$ patches which have roughly equal area, and discard in turn a different 
patch to calculate the CCF.
The estimator for the covariance matrix is in this case,
\bea \label {eq:covestJK}
{\cal {C}}_{ij} = \frac {M-1} {M} \sum_{k = 1}^{M}  &\left[ \hat C^{Tg}_{k} (\vartheta_i) - \bar C^{Tg} (\vartheta_i) \right]& \nonumber \\  \times 
 &\left[ \hat  C^{Tg}_{k} (\vartheta_j) - \bar C^{Tg} (\vartheta_j) \right],&
\eea
The advantage of this method is its model independence, but it has the big 
drawback of giving different answers depending on the size and 
number of the discarded areas.  It also implicitly assumes independence of the various patches, which is not always the case. 

Our ultimate goal is to measure the total covariance matrix between all the catalogues. 
To do so, we need to estimate the total covariance matrix $ {\cal{C}}_{ij}^{pq} $, as the 
matrix that has in the diagonal blocks the single catalogue $ {\cal {C}}_{ij}^{pp} $, and in 
the off-diagonal parts is 
\be \label {eq:covestMCtot}
{\cal {C}}_{ij}^{pq} = \frac {1} {M} \sum_{k = 1}^{M} \left[ \hat C^{Tp}_{k} (\vartheta_i) - \bar C^{Tp} (\vartheta_i) \right] \left[ \hat C^{Tq}_{k} (\vartheta_j) - \bar C^{Tq} (\vartheta_j) \right].
\ee
For simplicity, we redefine the indexes $ i,j $ in a way that they run from $ 1 $ to 
$ N_{\mathrm{tot}} = N_{\mathrm{bin}} \times N_{\mathrm{cat}} $, i.e. redefining the 
data, theory and mock arrays as the concatenation of all catalogues' CCFs with the CMB.  In this way,
the covariance matrix is simply the square matrix ${\cal {C}}_{ij}$,  
identical to the Eq.~(\ref {eq:covestMCtot}) but now with dimension 
$ N_{\mathrm{tot}} $. A similar expression can be easily obtained for the JK case.

\section {The Catalogues} \label {sec:cats}

To best detect the ISW effect through the cross-correlation technique, we ideally require 
surveys covering large fractions of the sky, so that accidental correlations will cancel out. 
The surveys also need to be sufficiently deep, in order to probe the gravitational potentials 
where the ISW effect is being created.  Ideally, we would like to span the redshift range 
$ 0 < z < 3 $, separated into subsamples of different depths so as to measure the redshift dependence of the 
effect and get some handle on the evolution of the dark energy.   However, only rather coarse redshift 
information is required, so redshift errors of $\Delta z \sim 0.1$ obtainable through photometric methods should be sufficient for these purposes. 
This is beyond the present state of the observations, but the differences in the redshift distributions of the various samples 
does provide some limited information on the dark energy evolution. 

At present, the best surveys available for this purpose (and where ISW detections have previously been found) include  
the following: the optical Sloan Digital Sky Survey (SDSS), the infrared 2 Micron All-Sky Survey (2MASS), the X-ray 
catalogue from the High Energy Astrophysical Observatory (HEAO) and radio galaxy catalogue from the NRAO VLA Sky Survey (NVSS). 
The high quality of the SDSS data allows us to extract some further subsamples from it, consisting of 
Luminous Red Galaxies (LRG) and quasars (QSO) in addition to the main galaxy
sample \cite{Peiris:2000kb}.  These are the samples we use in our analysis below, and include most of the significant 
reports of the ISW detection.  Because the data has not been publicly released and since it is not significantly deeper than 2MASS, we omit the APM galaxy survey, which has also been reported to have evidence for ISW cross-correlations \cite{Fosalba:2003iy}.  

\begin{figure}[ht] 
\begin{center}
\includegraphics[angle=0,width=1.0\linewidth]{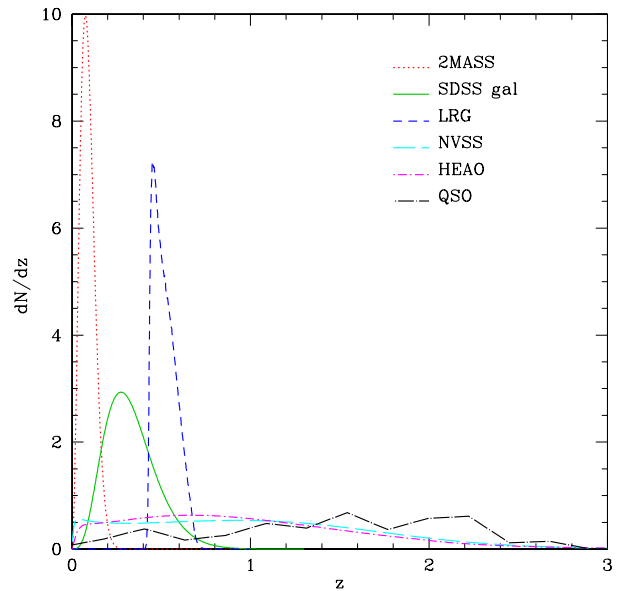}
\caption{The redshift distributions of all catalogues $ dN/dz $ normalised to 
unity.  The significant overlap between redshift distributions (especially for the X-ray and radio surveys) results in a  
covariance matrix with significant non-diagonal elements.}
\label {fig:dndz-all}
\end{center}
\end{figure}

We show in Fig.~\ref{fig:dndz-all} the redshift distributions $ dN/dz $ of 
the catalogues we use, normalised to unity; we can see that they span a redshift 
range $ 0 < z < 2.5 $, similar to the theoretical requirement, although the 
overlap between different samples is significant. This 
means that the covariance between the measures could be large: one of 
the goals of this paper is to quantify it.

In the rest of this section, we will present the characteristics of all 
the samples we use, in order of increasing redshift.

\subsection{2MASS}

The 2 Micron All Sky Survey (2MASS) is an infrared catalogue; its extended 
source catalogue (XSC)~\cite{Jarrett:2000me} contains $\sim 800,000$ galaxies 
with median redshift $ z \sim 0.1 $ and, unlike the point source catalogue (PSC) is almost free of stellar contamination.  Some evidence for ISW cross-correlations
has been seen in 2MASS previously \cite{Afshordi:2003xu, Rassat:2006kq}, and we largely follow 
the galaxy selection of those previous analyses here.   

Accordingly, we select galaxies according 
to their $ K_s $-band isophotal magnitude $ K_{20} $ (\texttt{k\_m\_i\_20\_c}, 
- 20 mag / arcsec$^2$). These magnitudes are corrected for Galactic 
extinction using the infrared reddening maps by~\cite{Schlegel:1997yv}, 
as $ K_{20}' = K_{20} - A_K $, where the extinction is $ A_K = 0.367 (B-V) $.
The requirement of completeness of the catalogue is satisfied by imposing 
a cut in magnitude $ K_{20}' < 14.0 $, while we can exclude low redshift 
sources with the condition $  K_{20}' > 12.0 $.
We only include objects with a uniform detection threshold ($\mathtt{use\_src} = 1$), and remove known artifacts  
($ \mathtt{cc\_flag} \neq  \mathtt{a}  $ 
and $  \mathtt{cc\_flag} \neq \mathtt{z} $); we 
also exclude a small fraction of objects where the magnitude or its error were not recorded.  

In addition to the pixelisation geometry mask, we follow earlier analyses 
\cite{Afshordi:2003xu, Rassat:2006kq} excluding areas of the sky with high 
reddening, discarding pixels with $ A_k > 0.05 $; this leaves $ 69 \% $ of 
the sky and 718,000 galaxies after excluding artifacts.
It is reported by~\cite{Afshordi:2003xu, Rassat:2006kq} that the redshift distribution 
of these galaxies is well approximated by the function:
\be \label {eq:dndz_std}
\frac{dN} {dz} = \frac {1} {\Gamma \left( \frac {m + 1} {\beta} \right)} \beta \frac {z^m} {z_0^{m+1}} \exp {\left[ - \left( \frac {z} {z_0} \right)^{\beta} \right]}
\ee
where the parameters are  $ z_0 = 0.072 $, $ \beta = 1.752 $ and $ m = 1.901 $.
This distribution is shown together with the others in Fig.~\ref {fig:dndz-all}.

To check the consistency of the dataset and its bias we calculate its 
auto-correlation function (ACF). The measure is in good agreement with the 
predictions for the best fit \WMAP model with a galactic bias $ b_g = 1.4 $ 
as found by \cite {Rassat:2006kq}, as we can see in Fig.~\ref {fig:aball}.

\subsection{SDSS galaxies}

The SDSS Sixth Data Release (DR6)~\cite{:2007wu, York:2000gk} is the largest wide optical 
galaxy survey available at the present for the northern hemisphere. From this 
catalogue we select a magnitude limited subsample $ 18 < r^{\star} < 21 $; this catalogue contains 30 million galaxies.
Here $ r^{\star} $ is the extinction corrected $r$ SDSS \"ubercalibrated model 
magnitude, i.e. using the SDSS variables $ r^{\star} = \mathtt {ubercal.modelMag\_r}  - \mathtt {extinction\_r}$: this corresponds to the procedure of 
\cite{Cabre:2006qm}, with the difference of using the sixth data release 
and the \"ubercalibrated model magnitude instead of the Petrosian magnitude, 
which is less reliable for faint objects.
We apply the pixelisation geometry mask and, in addition, 
we discard the pixels most affected by reddening, with $ A_r > 0.18 $. We 
also discard the southern stripes, since they are most affected by foregrounds and edge effects.

We select only objects with photometric redshifts between  
$ 0.1 < z < 0.9 $ and with an error on the redshift $ \sigma_z < 0.5 z $, leaving  $ 23.5 $ million galaxies in the catalogue. 
We could use these photometric redshifts as the basis of the theoretical calculations; however, since the distribution of the 
photometric redshifts can be affected by singularities in the redshift determination procedure, we use instead a fit to their distribution 
with the smooth function of Eq.~(\ref {eq:dndz_std}). The best fit parameters 
are in this case  $ z_0 = 0.113 $, $ \beta = 1.197 $ and $ m = 3.457 $, 
corresponding to a median redshift $ z_{\mathrm{med}} = 0.32 $.  (The results are actually independent of whether the fit or the 
actual redshift distribution is used).  The fit is shown 
together with the others in Fig.~\ref {fig:dndz-all}.

The ACF is in agreement with the prediction for the \WMAP best fit cosmology 
and a bias $ b_g = 1 $, as we can see in Fig.~\ref {fig:aball}.

\subsection{SDSS LRG}

Luminous Red Galaxies (LRGs) from the SDSS have been used often to find evidence for the 
ISW effect, as they have a deeper redshift distribution than the ordinary galaxies, with a mean 
redshift of $z \sim 0.5$~\cite{Fosalba:2003ge,Scranton:2003in,Padmanabhan:2004fy}.
In this analysis we use the MegaZ LRG sample~\cite{Collister:2006qg,Blake:2006kv} 
which contains 1.5 million objects from the SDSS DR6 selected with a 
neural network \cite{MegaZ}.  To ensure completeness we require that $i<20$.   To reduce 
stellar contamination we implement cuts on $ \delta_{sg} $, which is a variable of the MegaZ neural network estimator, defined such that $ \delta_{sg}  = 1 $ if the  object is a galaxy, and $ \delta_{sg}  = 0 $ if it is a star \cite{Collister:2006qg}. Following the conservative suggestion by \cite{Collister:2006qg}, we choose a cut  $ \delta_{sg}  > 0.2 $, which is reported to reduce stellar contamination below 2\% while keeping 99.9 \% of the galaxies. Stricter cuts have been tried with no significant changes to the CCF.

The mask we apply to this catalogue is a combination of the pixel geometry 
mask and two foreground masks, to account for seeing (cutting pixels with 
median seeing in the red band greater than 1.4 arcsec) and reddening (cutting pixels 
with median extinction in the red band $A_r > 0.18$).
The redshift distribution function in this case is found directly from the photometric 
redshifts that are given in the catalogue, and is shown in Fig.~\ref {fig:dndz-all}.

We show the auto-correlation function in Fig.~\ref{fig:aball}, where we can 
see that this is in agreement with the theoretical prediction from the best 
fit \WMAP cosmology and a bias $ b_g = 1.8 $, which is compatible with the estimate $ b_g = 1.7 \pm 0.2 $ shown by \cite{Blake:2006kv}, 
although some excess power at large scales is present, which might be explained 
as being produced by a residual stellar contamination.

\subsection{NVSS}

The NRAO VLA Sky Survey (NVSS) is a flux limited radio survey at a frequency 
of 1.4 GHz, with a minimum flux of $\sim 2.5$ mJy.  It is complete for declinations $ \delta > -40^{\circ} $, covering roughly 
80\% of the sky and contains $ 1.8 \cdot 10^6 $ sources. 
The mask to this catalogue is a combination of the most aggressive \WMAP mask 
(\emph{kp0}) plus a cut around point sources as described in 
\cite{Boughn:2001zs}, which also describes corrections made for a 
systematic in the mean density as a function of declination.    
The cross-correlations between NVSS and \WMAP have been observed by a number of groups, both in 
the correlation function \cite{Boughn:2003yz, Nolta:2003uy} and using an array of wavelet techniques
\cite{Vielva:2004zg,McEwen:2006my,Pietrobon:2006gh,McEwen:2007rz}.

The redshift distribution is uncertain; we base ours on models of Dunlop and Peacock \cite{Dunlop:1990kf} which 
seem to be still widely accepted, and are largely consistent with observations of cross-correlations with other surveys (though see 
below for further discussion).  
We calculate the auto-correlation function and present it in 
Fig.~\ref{fig:aball}; there is good agreement with the theory from the 
\WMAP best fit model and a galactic bias $ b_g = 1.5 $, compatible with the result $b_g = 1.5 \pm 0.2 $ by \cite{Boughn:2001zs}, although we see some 
excess power at small scales.

\subsection{HEAO}

The High Energy Astrophysical Observatory (HEAO1-A2) data set is a full
sky flux map of hard X-rays counts in the $2-10$ keV energy range
\cite{Boldt:1980cd}. We use the map and the mask determined
by~\cite{Boughn:1997vs,Boughn:2002bs}: the map is masked for the galactic
plane, a round area around the galactic centre and patch areas around
bright point sources.  The redshift distribution is also uncertain and
provided by modelling the X-ray background, as described in
~\cite{Boughn:1997vs,Boughn:2004ah}.

The modelling of the theoretical ACF for this catalogue is more complex
than those considered above, in that we are looking at flux rather than
number counts and the experimental beam is large compared to the pixel
size.  (The point spread function of the beam is well modelled by a
Gaussian with a full width, half maximum size of
$\vartheta_{\mathrm{FWHM}} = 3.04^{\circ} $ \cite{Boughn:2002bs}).  In
addition, the number of photons is small, so there is an additional
contribution from the photon shot noise.  Thus, the observed correlation
is the sum of three terms: the intrinsic correlations, the Poisson
correlations due to finite numbers of sources and shot noise due to the
finite number of photons.  The variance of the X-ray map is dominated by
photon shot noise (41\%) and Poisson correlations (45\%) while intrinsic
correlations are relatively small (14\%).  However, the shot noise
contributes only to the $0^\circ$ ACF while the Poisson correlations fall
off more quickly with angle than intrinsic correlations and become
sub-dominant for $\theta > 4^\circ$.  Consequently, the combination of shot
noise and Poisson correlations are not the primary component of the
total noise in the ISW signal.
We can see in Fig.~\ref {fig:aball} that the total modelled ACF 
fits the observations on large angles, assuming the \WMAP best fit model and a galactic 
bias $ b_g = 1.06 $, as found by \cite{Boughn:2003yz}.

\subsection{SDSS QSO}

The quasar survey we use comes from the SDSS DR6 through the NBC-KDE catalogue by 
\cite{Richards:2004cz, Richards:2008}, that contains over a million quasars.  
This new DR6 edition of the catalogue does not include as many parameter cuts as did
 the previous DR4 version. To obtain the cleanest possible dataset, and for
 consistency with our previous measure of the cross-correlation~\cite{Giannantonio:2006du},
 we only used quasar candidates selected via the UVX-only criteria used in the
 previous version of this photometric quasar catalogue. In addition, we consider only objects with a \emph {good} (positive) quality flag.
Following our previous results~\cite{Giannantonio:2006du} we impose a cut 
in reddening, discarding areas with $ A_g > 0.18 $. After these cuts, we are left with $N \simeq 500,000 $ quasars.

\begin{figure*}[ht] 
\begin{center}
\includegraphics[angle=0,width=1.\linewidth]{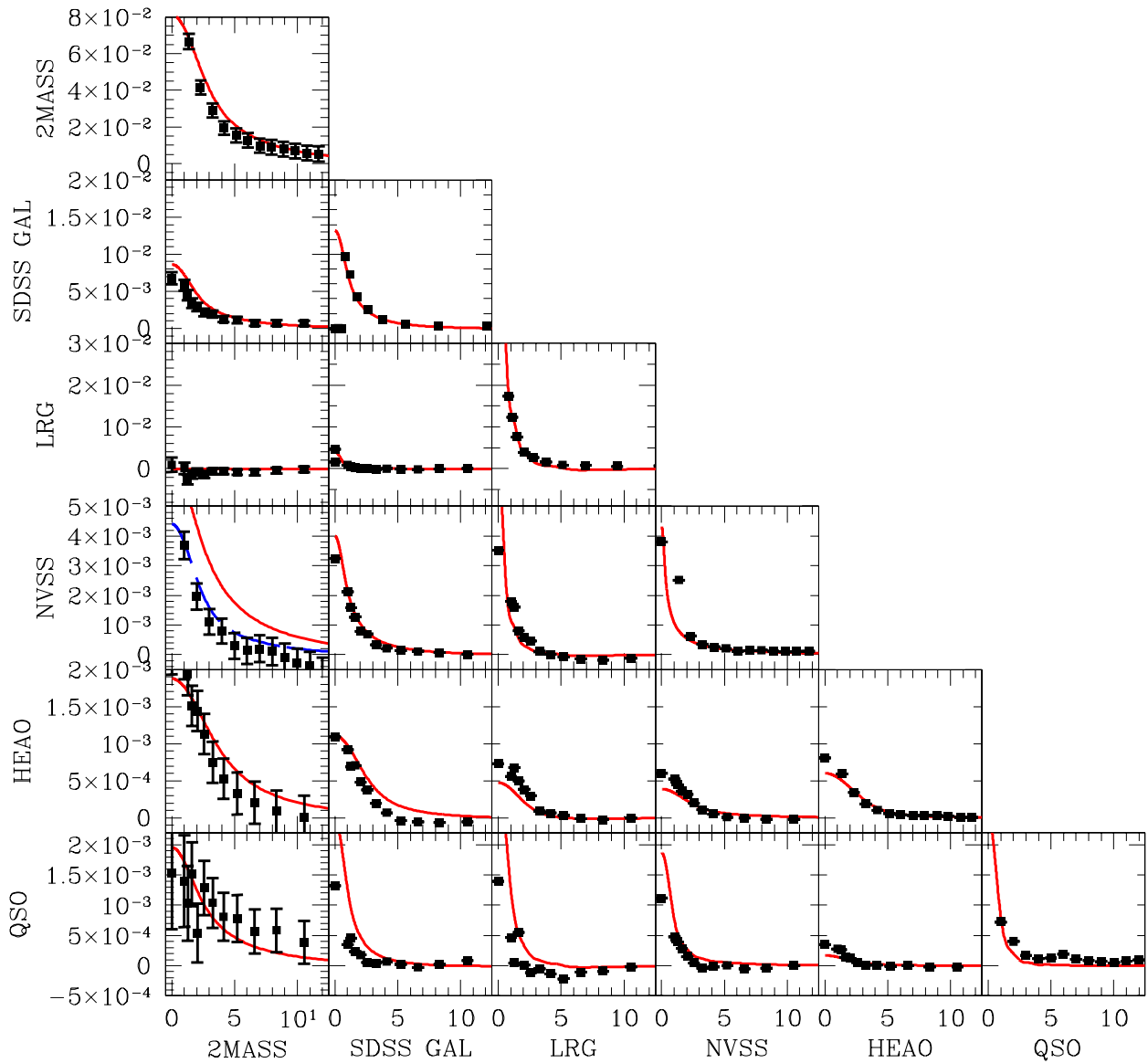}
\caption{Measures of the two-point correlation functions between all the 
combinations of catalogues, where the units in the x-axis are degrees.  The auto-correlations are on the diagonal, 
and the solid (red) lines show the theory from \WMAP best fit cosmology and the galactic bias 
from the literature.  The largest discrepancy with theory, in the NVSS-2MASS CCF, can be addressed by a small change in the assumed
NVSS redshift distribution (blue dashed line).}
\label {fig:aball}
\end{center}
\end{figure*}

This catalogue comes with estimated photometric redshifts, upon which we base 
the redshift distribution shown in Fig.~\ref{fig:dndz-all}.
There is evidence of some excess power in the ACF on large angular separations that indicate faint stars 
are still present in the catalogue after these cuts, as seen before in \cite{Giannantonio:2006du}. The amount of stellar contamination is $ \sim 3 \% $, as found by \cite {Richards:2008}, from comparison with the ACF of a random sample of stars taken from the SDSS, and does not contribute to the correlation with the CMB, as expected.
We can see in Fig.~\ref {fig:aball} the ACF for this sample; this is in good agreement with theoretical expectations and determines the bias of $b_g = 2.3$, as previously found in \cite {Myers:2005jk,Giannantonio:2006du}.

\section {Results} \label {sec:results}

In this section we present the measurements of the all the correlation functions
between the data sets we consider and their covariance. 

\subsection {Density-density cross-correlations} \label {sec:acf}

We begin by examining the cross-correlations between the different density maps. 
These measurements are shown in Fig.~\ref {fig:aball}, with the auto-correlation 
measurements along the diagonal.  This is the first measurement of the cross-correlations between 
most of these data sets.  The error bars are estimated by Monte Carlo realisations of all the data sets (MC2, as 
described above). 

The measurements largely agree with their theoretical predictions, which are based on the \WMAP best fit 
model using the visibility functions in Fig.~\ref{fig:dndz-all} and a linear bias for each.  The agreement is to be 
expected for the auto-correlations, which were the basis for the estimates of the linear bias.  
However, the cross-correlation measurements provide a useful consistency check for our model, and 
in particular for the visibility functions, since the cross-correlations are most sensitive to the degree that 
the measurements overlap in redshift.   

The largest discrepancy between the measurements and theory is in the NVSS-2MASS cross-correlation, where the 
theory is roughly twice as large as expected.   This is perhaps not unexpected, since the NVSS visibility function is known to 
be uncertain, and the overlap with 2MASS is in a narrow region of redshift.  It does indicate that less of the NVSS correlations are arising from the 2MASS redshift range  than expected in the model.  This could be because either the low redshift tail of the NVSS visibility function is 
overestimated relative to the high redshift region, or because the bias of the radio galaxies increases as we move to higher redshift.  
This can be addressed by a small change in the visibility function, as demonstrated by the blue dashed line in the panel (in this case we 
arbitrarily imposed a low redshift exponential damping in the visibility function, leaving the rest unchanged).   Such a change does not 
significantly affect the expected CMB cross-correlations considered here. 

\subsection {Temperature-density cross-correlations} \label {sec:ccf}

\begin{figure*}[htp] 
\begin{center}
\includegraphics[angle=0,width=.9\linewidth]{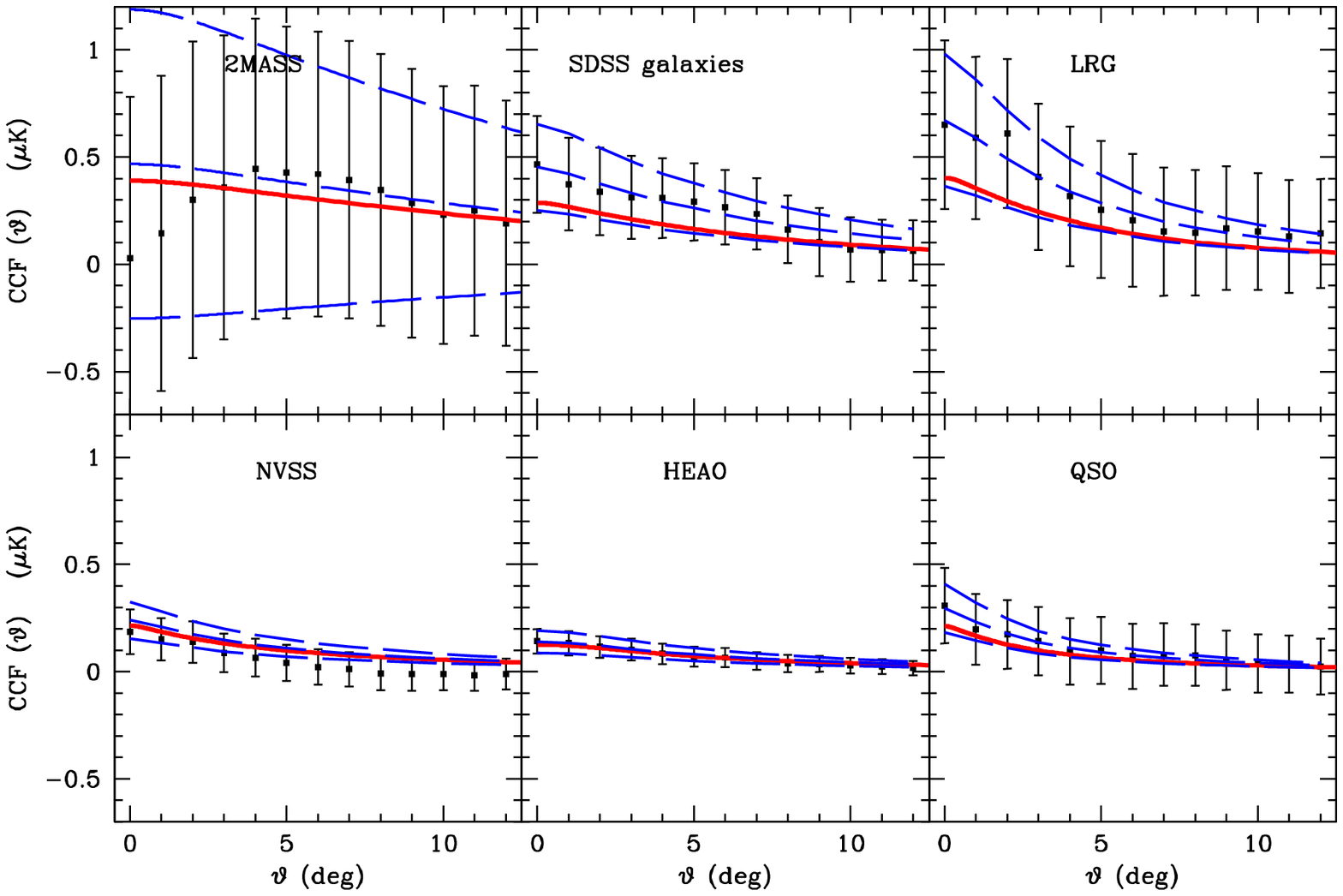}
\includegraphics[angle=0,width=.9\linewidth]{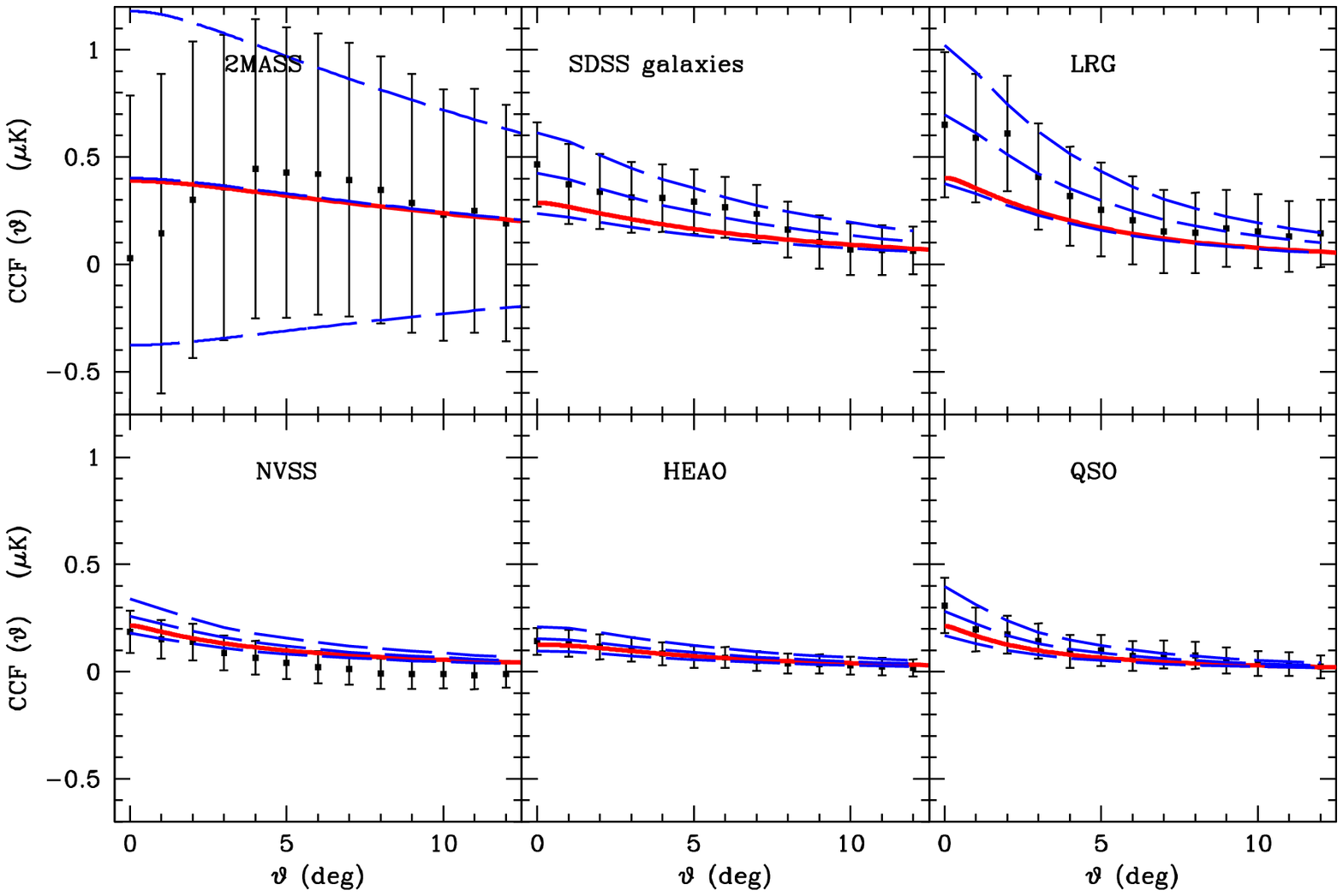}
\caption{Monte Carlo error estimation.  Measurements of the cross-correlation functions between all the catalogues and the \WMAP CMB maps (black points), compared with the theory from \WMAP best fit cosmology and the galactic bias from the literature (red solid lines). The best fit amplitudes and their $1-\sigma$ deviations are shown in blue (dashed).  
In the top panel, the errors are calculated with 5000 temperature-only Monte Carlos and, in the bottom panel, Monte Carlos for temperature and density including expected correlations.  We see that the errors are comparable for individual observations.  Because of known contamination from the Sunyaev-Zeldovich effect in the 2MASS data~\cite{Afshordi:2003xu}, the four smallest angle bins were excluded from the fits.}
\label {fig:ccf-all}
\end{center}
\end{figure*}

We next turn our attention to the cross-correlation functions 
between each density map and the CMB maps from \WMAP 3.
We use the \emph {internal linear combination} (ILC) maps from \WMAP, which 
are the cleanest data, although we have checked that the results do not 
depend on the frequency (see below), and we also apply the \emph {kp0} 
mask to them, cutting the galactic plane region. 
As we can see in Fig.~\ref {fig:ccf-all}, the measures are again largely in 
agreement with the theoretical predictions for the \WMAP best fit model. 

We have also checked the results obtained with the new \WMAP5 data, and we have not found  any difference in the correlations. This is expected, since \WMAP maps are already cosmic variance limited at large scales.

\begin{figure*}[ht] 
\begin{center}
\includegraphics[angle=0,width=0.9\linewidth]{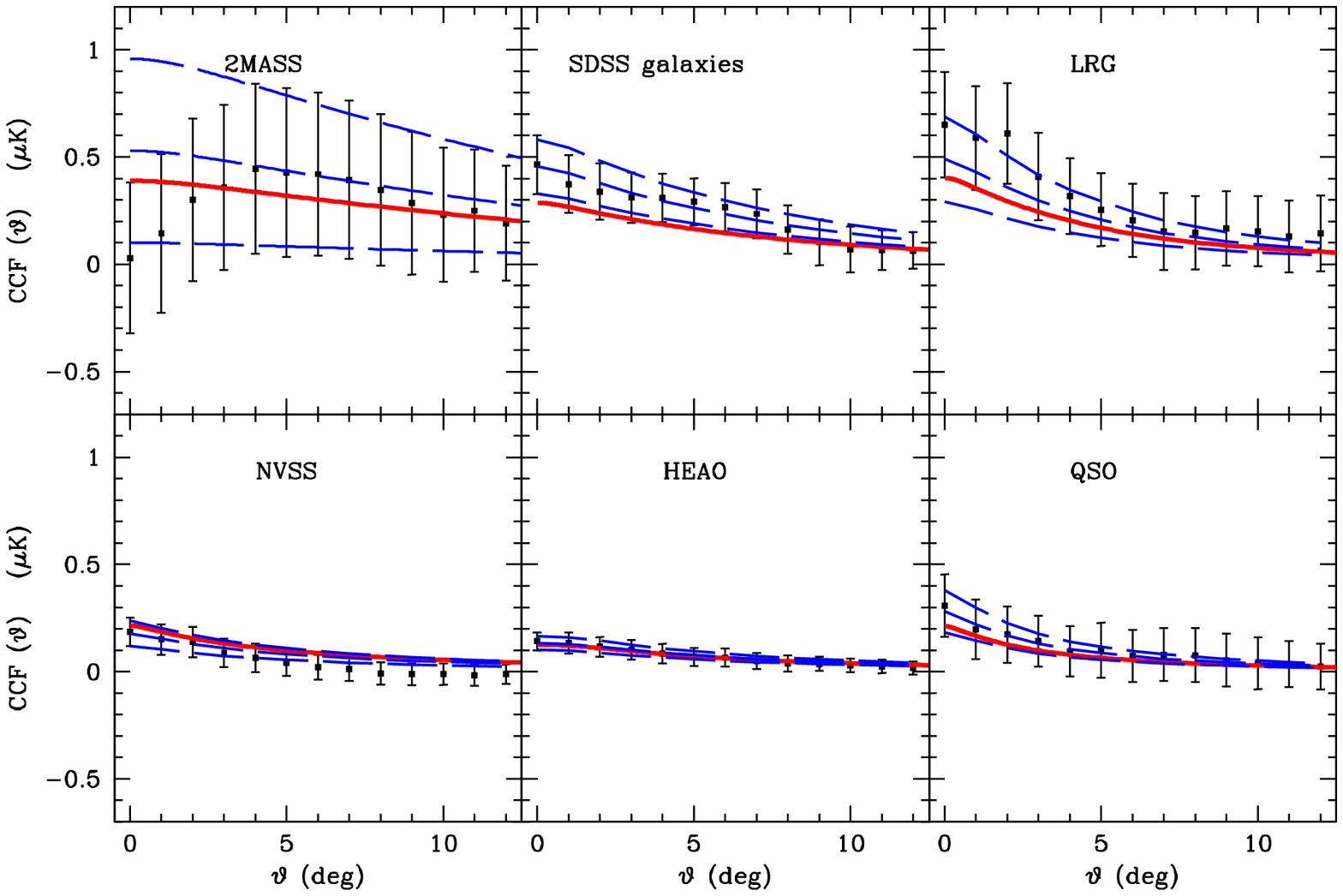}

\caption{Jack-knife error estimation. The lines are the same as in Fig.~\ref {fig:ccf-all}. The errors are somewhat smaller 
than seen from the Monte Carlo estimates, possibly due to correlations between the jack-knife subsamples. }
\label {fig:ccf-all-JK}
\end{center}
\end{figure*}

We now discuss the results obtained following the three methods of error 
estimation discussed in \S\ref{sec:theo} above.

\subsubsection {Temperature-only Monte Carlo errors}

We generate 5000 Monte Carlo simulations of the CMB anisotropy map with the 
\WMAP best fit parameters. We estimate the covariance matrix for each catalogue 
using Eq.~(\ref {eq:covestMC}), and the total covariance matrix follows from 
its generalisation.

\begin{figure}[ht] 
\begin{center}
\includegraphics[angle=-90,width=1.\linewidth]{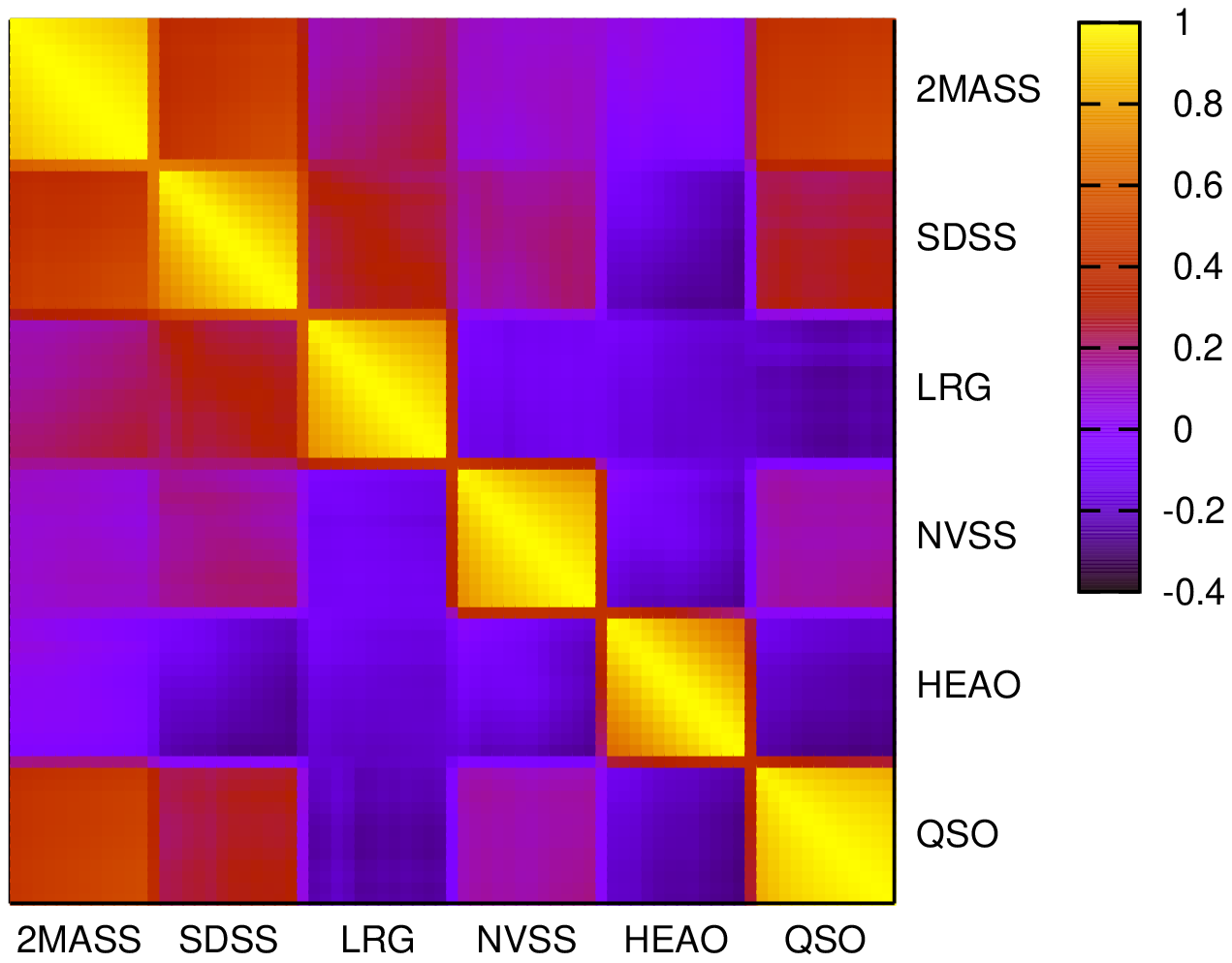}
\includegraphics[angle=-90,width=1.\linewidth]{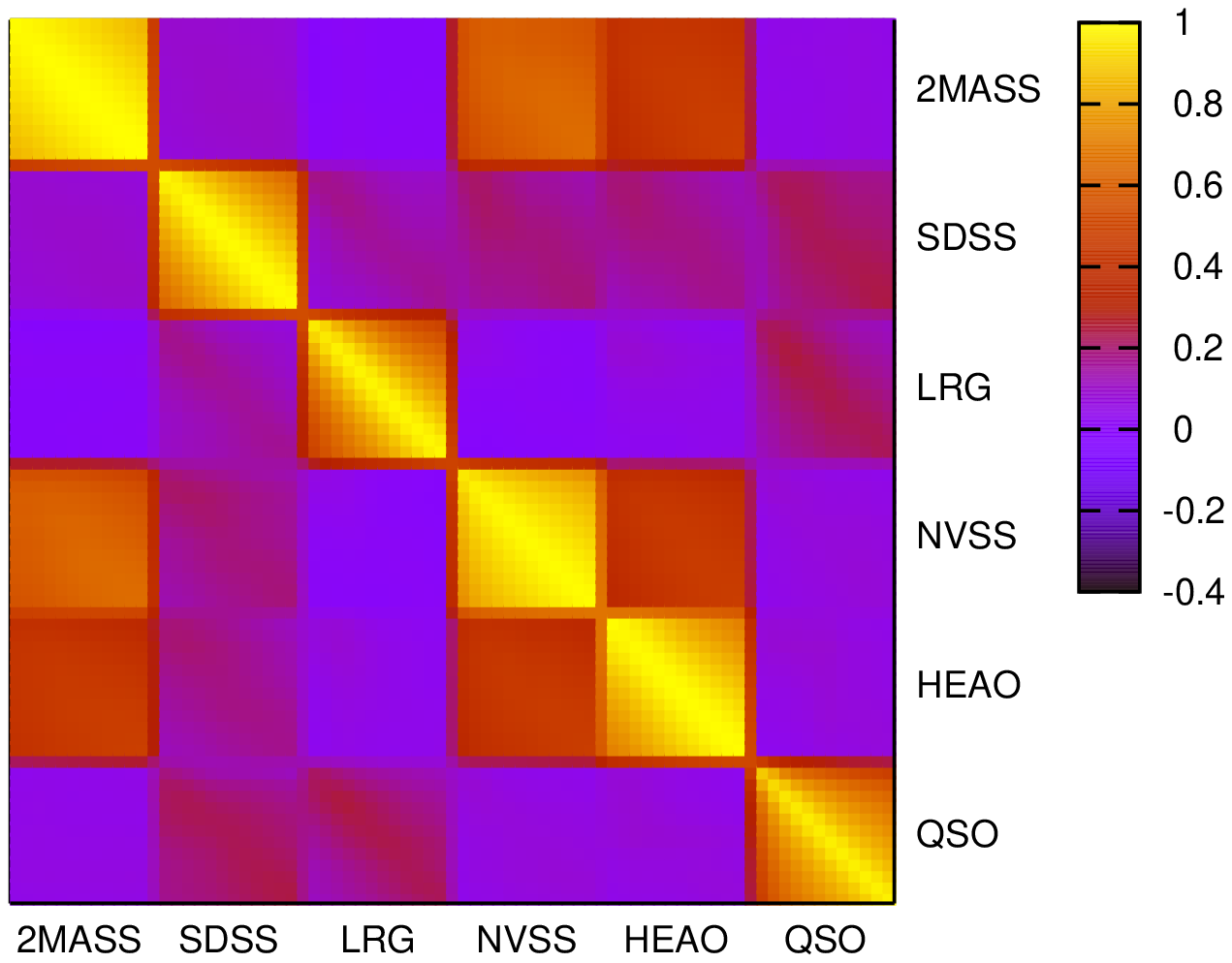}
\caption{The total covariance matrix obtained with 5000 Monte Carlos, normalised. The top panel shows the temperature-only Monte Carlos, while the bottom panel is the result of the full Monte Carlos. While the diagonal (single experiment) covariances are similar, those between 
experiments (off-diagonal) are somewhat different.}
\label {fig:covm}
\end{center}
\end{figure}

These are the errors shown in the top panel of Fig.~\ref {fig:ccf-all}; as 
we can see, the errors are quite large, especially for the low redshift 
catalogues, and the significance is further decreased by the high correlation 
between the points. 
We have checked that these errors converge; the convergence is already good 
after $ \sim 700 $ Monte Carlos for each single catalogue, and after 
$ \sim 3000 $ Monte Carlos for the full covariance matrix.  The covariance 
between the points is shown in Fig.~\ref {fig:covm}.

\subsubsection {Full Monte Carlo errors}

In this case, in addition to 5000 new mock CMB maps, we also generate 5000 mock 
density maps for each catalogue, correlated as expected theoretically, based on the \WMAP best cosmology and their 
redshift distributions.  In addition, the Poisson 
noise is added due to the expected number of objects per pixel.

The result calculated in this way is shown in the bottom panel of 
Fig.~\ref {fig:ccf-all}, and the relative full covariance matrix in the 
bottom panel of Fig.~\ref {fig:covm}. We can see that the errors estimated
in this way are generally consistent with their MC1 counterparts.

The largest difference between the approaches is in the covariance between the cross-correlations measured with 
different data sets (Fig.~\ref {fig:covm}).  Using the observed density maps yields both positive and negative covariance, while 
the covariance is only positive when all the maps are simulated.  In the first approach, the strongest correlations are between the 
SDSS subsamples and 2MASS.   In the second approach, which is purely theoretical, the largest covariances are between 2MASS, NVSS and HEAO.   The NVSS-HEAO covariance is expected to be large, since they are both essentially all sky maps and have similar redshift coverage. 
The large covariance between 2MASS-HEAO and 2MASS-NVSS is more surprising given the differences in the redshift distributions, but seem 
to be driven by the low redshift tail of the NVSS and HEAO distributions.  As noted above, the cross-correlations are smaller than expected theoretically for 2MASS-NVSS (and to a lesser extent for 2MASS-NVSS).  This indicates that the overlap of 2MASS with NVSS and HEAO is 
less than assumed, and that we have likely overestimated the covariance somewhat.  However, the low significance of the 2MASS CCF means this has a small impact on the final result.   
  
The differences between the two methods appear large for the off-diagonal elements.   The reasons for these differences are unclear,  
but they suggest that the observed density maps are somewhat atypical of those simulated.  
However, it is not surprising that any particular realisations would appear atypical in some way.   
Despite these differences, these covariance matrices give comparable final significance, as is discussed below. 

\subsubsection {Jack-knife errors}

For completeness, we also present the errors estimated from a jack-knife method.   
However, there is more ambiguity in implementing this method, leading to uncertainty in the 
resulting estimates in the errors.  

One issue is what patch size to use. Ideally for the jack-knife approach one would like the 
cross-correlation observations to be uncorrelated between patches.   In reality, some correlation is inevitable. 
We also need enough patches to estimate the full covariance matrix without it becoming singular, which drives the 
size of the patches down.   Thus, some kind of compromise is required. 

Since we are interested in the CCF on scales of a few degrees, we choose a patch size of order 10 square degrees. 
Because the surveys have different geometry and masks, the number of sectors $M$ will be different in each one.
The number of patches we can have in this way is generally low ($\sim 100$), so we cannot estimate the total 
covariance matrix which, having a dimension $ N_{\mathrm{tot}} = 78 $, requires 
at least a few hundred independent random measures to be correctly estimated.   

Cross-correlation measurements also introduce other issues, since the CMB and density maps are often covering different 
regions of the sky. 

In the end, we tried to be conservative and ensure the most independence between the subsamples by only including data
which were in the CMB and the density maps, and masking out both maps in the jack-knife estimates.   
The results we obtained are shown in Table \ref {tab:allresult}, where we compare the results obtained with 
jack-knife of the density map only and of both density and temperature maps using identical masks.

The jack-knife ambiguities are even more problematic when calculating the full covariance between observations 
using different density maps, since the density maps often will not overlap on the sky.   For this reason, and because such a large 
number of jack-knifes are required to estimate the total covariance matrix, we do not attempt to estimate it here. 

The error bars we estimate using the jack-knife method are of the same order of magnitude as those seen in the 
Monte Carlo approaches, but are somewhat smaller leading to higher significance in the detection.   This could be due to the lack of independence of the jack-knife patches, or because some aspect of the 
data is missing from the Monte Carlo approach.  We will use the Monte Carlo estimates below, focusing 
primarily on the results from MC2.

\subsection {Foregrounds \& systematics}

\begin{figure*}[ht] 
\begin{center}
\includegraphics[angle=0,width=0.9\linewidth]{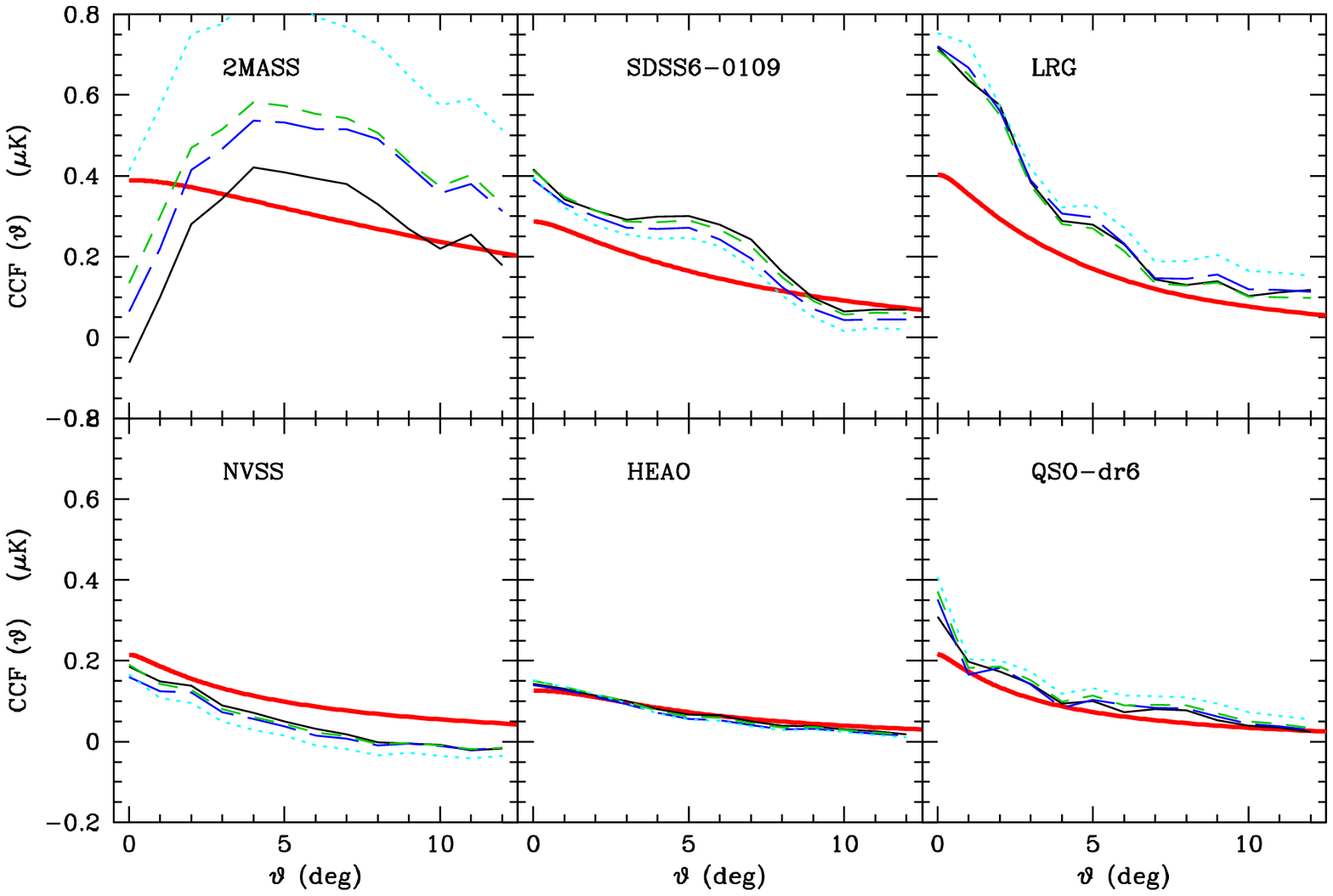}
\caption{Comparison of the CCF functions obtained with the different \WMAP frequency bands.
The black (solid) is using the internal linear combination map; the blue (long-dashed) uses the W band, the green (short-dashed) uses the 
V-band and the cyan (dotted) uses the Q-band.  The thick red curves show the \LCDM prediction.}
\label {fig:ccf-freq}
\end{center}
\end{figure*}

Since the ISW effect is gravitational in origin, it is frequency
independent as are the resulting CMB-density cross-correlations.  However,
a frequency dependence may in principle be introduced by foregrounds and
local contamination, such as the SZ effect. In Fig.~\ref {fig:ccf-freq}
we compare the CCF obtained with the different frequency bands from \WMAP
(ILC, W, V and Q bands), and we see that the result is substantially
independent of frequency, with the exception
of the 2MASS catalogue.   However, the 2MASS CCF detection is of low
significance and our final answers are not greatly sensitive to its inclusion.

Foreground contamination of the ISW signal is generally produced at low
redshifts.  A good way to make sure that such effects are not dominating
the measurement is to check for the sensitivity to the masking of these
foregrounds (e.g. \cite{Giannantonio:2006ij}).  For samples derived from
the SDSS (galaxies, LRGs and QSO), we test for foreground effects by
cutting the 20\% of pixels with the highest reddening (extinction),
seeing, sky brightness, and number of unresolved point sources. The most
relevant masks are the reddening and seeing masks which do not
substantially change the results.  For the other samples (2MASS, HEAO and
NVSS), we do not explore the masking, but we refer to the foreground
analyses presented in earlier papers ~\cite{Afshordi:2003xu,
Rassat:2006kq,Boughn:2001zs, Boughn:2002bs}.

\subsection {Comparison with previous measures}

We briefly compare our CCF measurements to others in the literature.

\subsubsection{2MASS}

From Fig.~\ref{fig:ccf-all} it is clear that the CCF for the 2MASS survey is consistent
with zero.  Previous analyses of these data found some evidence for
a positive correlation~\cite{Afshordi:2003xu, Rassat:2006kq}; however,
these were performed in Fourier space and included modelling of the SZ
effect, which manifests itself with anti-correlations at small angular
scales.  Indeed, it appears in Fig.~3 that the observed CCF turns over
at small angles.  If the smallest four angular bins are removed, the fit
to the CCF is consistent with the \LCDM theory; however, it is only
significant at the $\sim 1 \sigma$ level.  In any case, 2MASS appears to have
the least significant evidence for cross-correlations.

\subsubsection{SDSS galaxies} 

The main galaxy sample from the SDSS has a measured CCF which is also in 
good agreement with the theory. In this case, 
we note that we do not find agreement with the previous result of 
\cite{Cabre:2006qm}, who reported a measured CCF of almost double 
the amplitude that we detect. 

After discussions with the authors~\cite{Cabre:2006qm}, we  jointly
found this discrepancy resulted from an additional cleaning cut, where they discarded all galaxies 
with a large error on their Petrosian $r$ magnitude, imposing the condition 
$ \mathtt{petroMagErr\_r} < 0.2$. Imposing this same condition, we found that we could reproduce their result. 
Further, masking those areas with high proportion of Petrosian error also gave similar results. 

However, the motivation for such a cut is unclear. 
It is known that the Petrosian magnitudes are not accurate for faint objects, for which the best estimator 
is the \texttt{model} magnitude \cite{AdelmanMcCarthy:2007wh}.
 While having objects
with a well measured magnitude is desirable, we see no reason why cutting
galaxies on the basis of a poor estimate of their magnitudes should double
the correlation with the CMB.
This could happen if it were produced by some 
foreground mechanism, such as seeing or reddening, but we checked that none 
of the possible foreground maskings raised the CCF in any way comparable 
to the aforementioned cut. 

Therefore, lacking a valid reason to include this cut, and preferring to be conservative, we do not make the Petrosian error cut and 
our CCF is thus lower than seen by Cabr\'e et al. \cite{Cabre:2006qm}.  While it is worrying that a choice of masking has such a dramatic 
effect on the amplitude of the observed cross-correlation, it should be noted that the cross-correlation was largely independent of other masking choices.

\subsubsection{SDSS MegaZ LRGs}

The result for the LRG is the highest in comparison with the \LCDM theory.  
It agrees with the result of~\cite{Cabre:2006qm}.  A direct comparison with 
\cite {Scranton:2003in} and~\cite{Padmanabhan:2004fy} is more difficult
because these analyses use multiple photometric redshift bins.  Concentrating
on~\cite {Scranton:2003in} (since it also does its analysis in physical 
space, rather than Fourier space), we find approximately the same detection
significance as their single redshift bin measurements for similar data sets.
An updated version of this paper (available on the astro-ph archive, but also
unpublished) calculates a global $\chi^2$ value using all four of their LRG
samples, and detects a CCF with significance somewhat higher than we measure
in this work.  This is likely due in part to a somewhat larger redshift
baseline for their measurement as well as the fact that they calculated their
covariance matrix using a method similar to our MC1 case.  As one can see from
Fig.~\ref{fig:covm}, samples which cover very similar areas and
have significant redshift overlap (as is the case with their LRG photometric
redshift samples) can result in stronger anti-correlation between samples than
one observes in covariance matrices generated with the MC2 method.  This, in 
turn, would lead to a moderate over-estimation of the detection significance.

\subsubsection{Other measurements}

Not surprisingly, since we use the same maps generated from HEAO and NVSS, our results are in agreement with 
previous measures by~\cite{Boughn:2003yz, Nolta:2003uy}, and the amplitudes are consistent with the theoretical predictions.
As discussed above, the new Monte Carlo approach give consistent answers for individual experiments as the temperature-only
Monte Carlo approach used in earlier analyses.

We found that the measured 
CCF for the quasars is consistent with the earlier measurement and the expectation from theory, and it is independent from the cleaning level of the catalogue.    

In conclusion, all the measured CCF agree with the previous results and 
with the ISW theory for a \LCDM model, although they are in some cases marginally  
higher than theory predicts.

\section {Significance of the Result} \label{sec:signif}

Having established the measures of the CCFs and the total covariance matrix, 
we discuss the significance of this result and its consequences.

\subsection {Single catalogue significance}

Assuming that the detected cross-correlations are due to the integrated 
Sachs-Wolfe effect, we can assign a significance value to the measure if 
the errors on the cross-correlation are taken to be Gaussian.   For each 
catalogue, we can compare the measured CMB-density cross-correlation $ \hat C (\vartheta_i) $ 
with the theoretical expectation obtained from the \WMAP best fit cosmological parameters 
with our modified version of the \texttt{cmbfast} code~\cite{Seljak:1996is}. 

We perform the likelihood analysis first described in~\cite{Boughn:1997vs}. 
The shape of the CCF for each catalogue is assumed to follow the \LCDM
predictions.  The theory template is
\be
\bar{C} (\vartheta_i) = A  g (\vartheta_i),
\ee
where $g (\vartheta_i) $ is theoretical prediction of the \WMAP best fit
model and $A$ is the fit amplitude, which will depend on the visibility
function of the catalogue in question.
Maximising the likelihood
\bea
\mathcal {L} & = & (2 \pi)^{-N / 2} [\det {\cal {C}}_{ij}]^{-1 / 2} \nonumber\\
& \times & \exp \left[- \sum_{ij} ({\cal {C}}_{ij})^{-1} (\hat C_i - \bar C_i) (\hat C_j - \bar C_j)/2 \right],
\eea
we can find the best value for each $A$,
\be
A = \frac {\sum_{i,j=1}^N {{\cal {C}}_{ij}^{-1} g_i \hat C_j } }{\sum_{i,j=1}^N {\cal {C}}_{ij}^{-1} g_i g_j},
\ee
and the variance
\be
{\sigma^2_A} = {\left[ \sum_{i,j=1}^N {\cal {C}}_{ij}^{-1} g_i g_j  \right]^{-1}}.
\ee
We can also simply obtain the signal to noise ratio as $ S/N = A / \sigma_A $.

The results obtained in this way with errors calculated with the three 
methods are summarised in Table \ref {tab:allresult}, and the resulting 
amplitudes and their errors can be seen in Fig.~\ref {fig:ccf-all} and 
Fig.~\ref {fig:ccf-all-JK}.  Here we have allowed a separate amplitude $A$ for each catalogue.
Note that while the observed CCF is the same for the different methods, differences in the covariance 
matrices can result in different best fit amplitudes.  

It is possible to check that the Monte Carlo estimation has converged after $N$ realisations by estimating the uncertainty on the errors. In detail, we use a jack-knife approach consisting in observing the effect of removing $ M = 10 $ different subsets of the $ N = 5000 $ realisations of the MC2 method. The estimator of the uncertainty on $S/N$ is
\be
{\sigma^2_{S/N}} = \frac {M - 1} {M} \sum_{i = 1}^{M} \left[ \left(S/N \right)_i - \overline {S/N} \right]^2,
\ee
where $\left(S/N \right)_i$ are the signal to noise ratios obtained with each subset of $ N-M $ Monte Carlos, and $\overline {S/N}$ is their average. We find in this way that the uncertainty on the $S/N$ is less than $ 5\% $, indicating the level to which our Monte Carlos have converged.

\begin {table*}
\begin {tabular}{| c |  c | c | c | c | c | c | c | c |}
\hline
           &  \multicolumn{2}{c|}{5000 T-only Monte Carlos}  &  \multicolumn{2}{c|}{5000 full Monte Carlos} &    \multicolumn{2}{c|}{ JK - $\delta$ only } & \multicolumn{2}{c|}{JK - $\delta $ and $T$} 
\\
\hline
 catalogue   & $A$ & S/N & $A$ & S/N & $A$ & S/N  & $A$ & S/N    \\
\hline
\hline
2MASS cut     
 & $ 1.22 \pm 1.87  $ & $ 0.7 \sigma $ & 
    $ 1.00 \pm 1.96 $ & $ 0.5 \sigma $  &  
    $ 0.66 \pm 0.77 $ & $ 0.9 \sigma $ &
    $ 1.36 \pm 1.10 $ & $ 1.2 \sigma $
\\
SDSS     
  & $ 1.58 \pm 0.70  $ & $ 2.2 \sigma $ &
    $ 1.48 \pm 0.66  $ & $ 2.2 \sigma $ &     
    $ 1.24 \pm 0.42  $ & $ 3.0 \sigma $ &
    $ 1.59 \pm 0.44  $ & $ 3.6 \sigma $         \\
LRG     
   & $ 1.67 \pm  0.76 $ & $ 2.2 \sigma $ &
   $ 1.73 \pm 0.80  $ & $ 2.2 \sigma $ & 
     $ 0.92 \pm 0.50  $ & $ 1.8 \sigma $ &
     $ 1.22 \pm 0.49  $ & $ 2.5 \sigma $        \\
NVSS    
   & $ 1.12 \pm  0.40 $ & $ 2.8 \sigma $ &
    $ 1.20 \pm 0.37  $ & $ 3.3 \sigma $ & 
     $ 0.68 \pm 0.29  $ & $ 2.4 \sigma $ &
     $ 0.83 \pm 0.27  $ & $ 3.1 \sigma $       \\
HEAO     
  & $ 1.10 \pm  0.41 $ & $ 2.7 \sigma $  &
     $ 1.22 \pm 0.45  $ & $ 2.7 \sigma $  &
     $ 0.97 \pm 0.26  $ & $ 3.7 \sigma $ &
     $ 1.00 \pm 0.24  $ & $ 4.2 \sigma $       \\  
QSO      
  & $ 1.40 \pm  0.53 $ & $ 2.6 \sigma $ &
        $ 1.33 \pm 0.54  $ & $ 2.5 \sigma $ &     
     $ 1.50 \pm 0.58  $ & $ 2.6 \sigma $ &
     $ 1.33 \pm 0.46  $ & $ 2.9 \sigma $       \\
\hline
\hline
\textbf{TOTAL} 
  &  $ \mathbf{1.02 \pm 0.23}  $ & $ \mathbf{4.4 \sigma} $ &
  $ \mathbf{1.24 \pm 0.27}  $ & $ \mathbf{4.5 \sigma} $    & --- & --- & --- & --- \\
\hline
\end {tabular}
\caption{The amplitudes and their significance for different methods of calculating the covariance.   The left columns show the two Monte Carlo methods, while the right two show the jack-knife method with equal area (10 deg$^2$), in one case masking only patches of the density map, and in the other masking both density and temperature maps.  We do not calculate the full  covariance matrix or the total significance for the jack-knife cases. For 2MASS, we have cut the first four angular bins because of their SZ contamination; the total significance is obtained discarding these bins.}
\label{tab:allresult}
\end{table*}

\subsection {Joint significance}

We can easily generalise this to combine the different catalogues and obtain a single significance. 
Redefining the indexes $ i,j $ in a way that they now run from $ 1 $ to 
$ N_{\mathrm{tot}} = N_{\mathrm{bin}} \times N_{\mathrm{cat}} $, running over each of the bins of the 
of the observed (and theoretical) cross-correlation functions for each of the density catalogues. 
Using the full covariance matrix, we can follow 
again the same procedure, and find a single best fit amplitude.   

The results obtained in this way are shown at the bottom of 
Table \ref{tab:allresult}. The significance of the two different Monte Carlo methods, MC1 and MC2, are $4.4 \sigma$ and $4.5 \sigma$ respectively.  We also find that the uncertainty on the $S/N$ for the joint amplitude is again less than $ 5\% $.

The two MC methods produce similar detection significances, but
this could be a lucky coincidence, since the covariance matrices
relating different surveys are much different.   
Both methods suggest some pairs of observations should be strongly
correlated, but which pairs are strongly correlated depends on the
method.  If the covariance between surveys were ignored, the total
significance would be about $5.8 \sigma$.  Perhaps it is not
surprising then that adding similar levels of covariance between
experiments with comparable individual detection levels would have a
similar effect on the total significance.

As in the case of fits to individual correlation functions, strong
covariances can have results which are counter-intuitive.  For example,
the fit for the total amplitude using the MC1 approach is smaller than
any single survey would suggest.   Also, adding the small angle 2MASS
CCF, believed to be suppressed by SZ, actually increases the fits by
about $0.2 \sigma$ despite the points themselves being lower than the
theory.  These effects suggest that the degree of covariance between the
different measurements might be over-estimated, which would not be
surprising given the much different systematics in each experiment.  
Even adding a small degree ($5\%$) of diagonal noise is enough to
increase the total MC1 amplitude to $A = 1.14 \pm 0.26$, with a corresponding $ S/N = 4.4 $, so that it is
more consistent with the amplitudes of the individual experiments.  The
MC2 result is not affected by such a change, because the total amplitude
is already consistent with the individual survey measurements.

Note that the theoretical model associated with a particular best fit amplitude is not unique.  While increasing the dark energy density will generally increase
the ISW effect, the effect will generally be redshift dependent and could impact different catalogues differently.  However, the \LCDM model without any tweaking ($A=1$) improves the likelihood at $ \sim 4.5 \sigma$ compared to the absence of cross-correlations.  Below we compare to specific alternative cosmologies without any scaling amplitude.

\subsection {$\chi^2$ Tests}

Another way to assess the significance of the measure with respect to 
a theory is simply to look at the $ \chi^2 $, defined as
\be
\chi^2 = \sum_{ij} {\cal {C}}^{-1}_{ij}  (\hat 
C_i - \bar C_i) (\hat C_j - \bar C_j),
\ee
where the inverse covariance matrix and the data can be referred either to a 
single catalogue or to the total measure.   Whereas the likelihood method 
discussed above looks at how well a model can reduce  $ \chi^2$, it is also 
worth simply looking at the magnitude of $ \chi^2$ for the null hypothesis test, 
where we calculate the $ \chi^2_0 $ assuming the theoretical cross-correlation is zero.

\begin {table}[h]
\begin {tabular}{|  c |  c  |  c  |  c  |  c  |}
\hline
 catalogue   &  $ f $  &  $ \:\:\:\: \chi^2_0 \:\:\:\: $  &  $ \chi^2_{\mathrm{best fit}} $  &  $ \chi^2_{\mathrm{\Lambda CDM}} $    \\
\hline
\hline
2MASS      &  9  &  5.4  &  5.2 &  5.2  \\
SDSS       &  13  &  17  &  11  &  12  \\ 
LRG        &  13  &  9.6  &  4.9  &  5.7  \\ 
NVSS       &  13  &  17  &  6.0  &  6.3  \\
HEAO       &  13  &  18  &  10  &  10  \\
QSO        &  13  &  9.7  &  3.7  & 4.0  \\
\hline
\hline
\textbf{TOTAL}  &  74  &  67  &  47  &  48  \\
\hline
\end {tabular}
\caption{A comparison of the absolute $\chi^2$ for the various experiments.}
\label{tab:allchi2}
\end{table}

In Table \ref {tab:allchi2}, we the show the $ \chi^2$ for the null hypothesis, as well as for 
the \LCDM and best-fit models.   We use the MC2 errors, dropping the first four bins 
of 2MASS which appear to be affected by SZ.   While there is much variation, 
in most cases there is not clear evidence against the null hypothesis, in that its $\chi^2_0$ is not 
significantly greater than the number of data points.   However, the $\chi^2$ values are significantly 
reduced if one assumes one of the models, like \LCDM, which predict a non-zero cross-correlation.

The reasons for the particularly low $\chi^2$ for the LRG case is unclear, and we investigate this more below.  It might be an indication
that the error estimates are in some sense too large, or that the covariance between angular bins is different than 
expected from the simple Monte Carlo simulations, perhaps as a result of foregrounds.   
However, it should be emphasised that the $\chi^2$ for the null hypothesis is fairly conservative, and unlike the 
Bayesian likelihood approach, it fails to account for the 
fact that we have strong theoretical expectations for the signal we are looking for.

\subsection {Eigenmode decomposition}
To better understand the covariance of our data, and especially to 
understand the $\chi^2$, it is useful to study the eigenmode  
decomposition of the covariance matrix.  As a worst-case example,  we will use here the measurement and covariance matrix for 
the LRG sample calculated with the MC2 method (dimension $n = 13$).

\begin{figure}[htp] 
\begin{center}
\includegraphics[angle=0,width=1.0\linewidth]{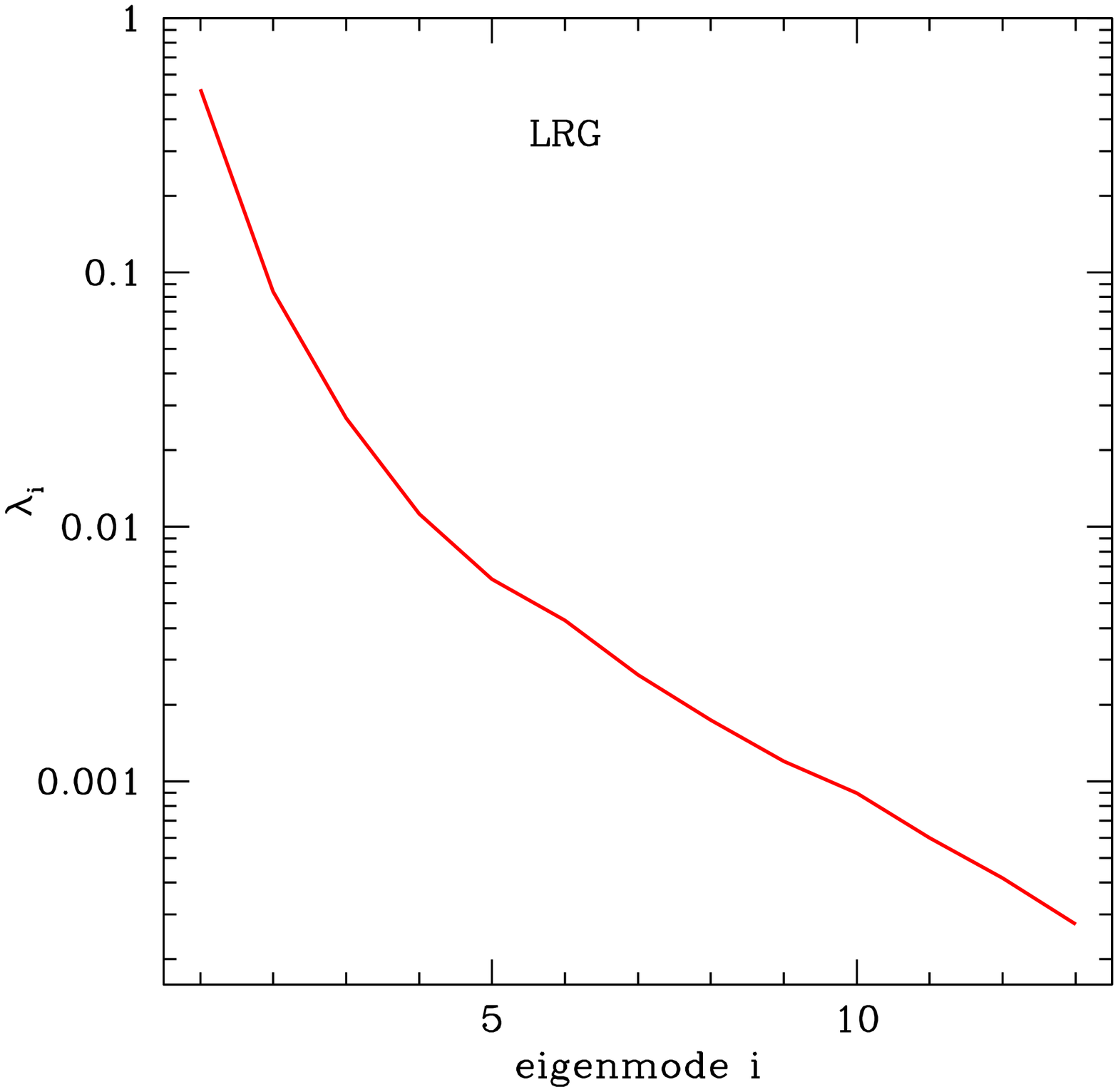}
\includegraphics[angle=0,width=1.0\linewidth]{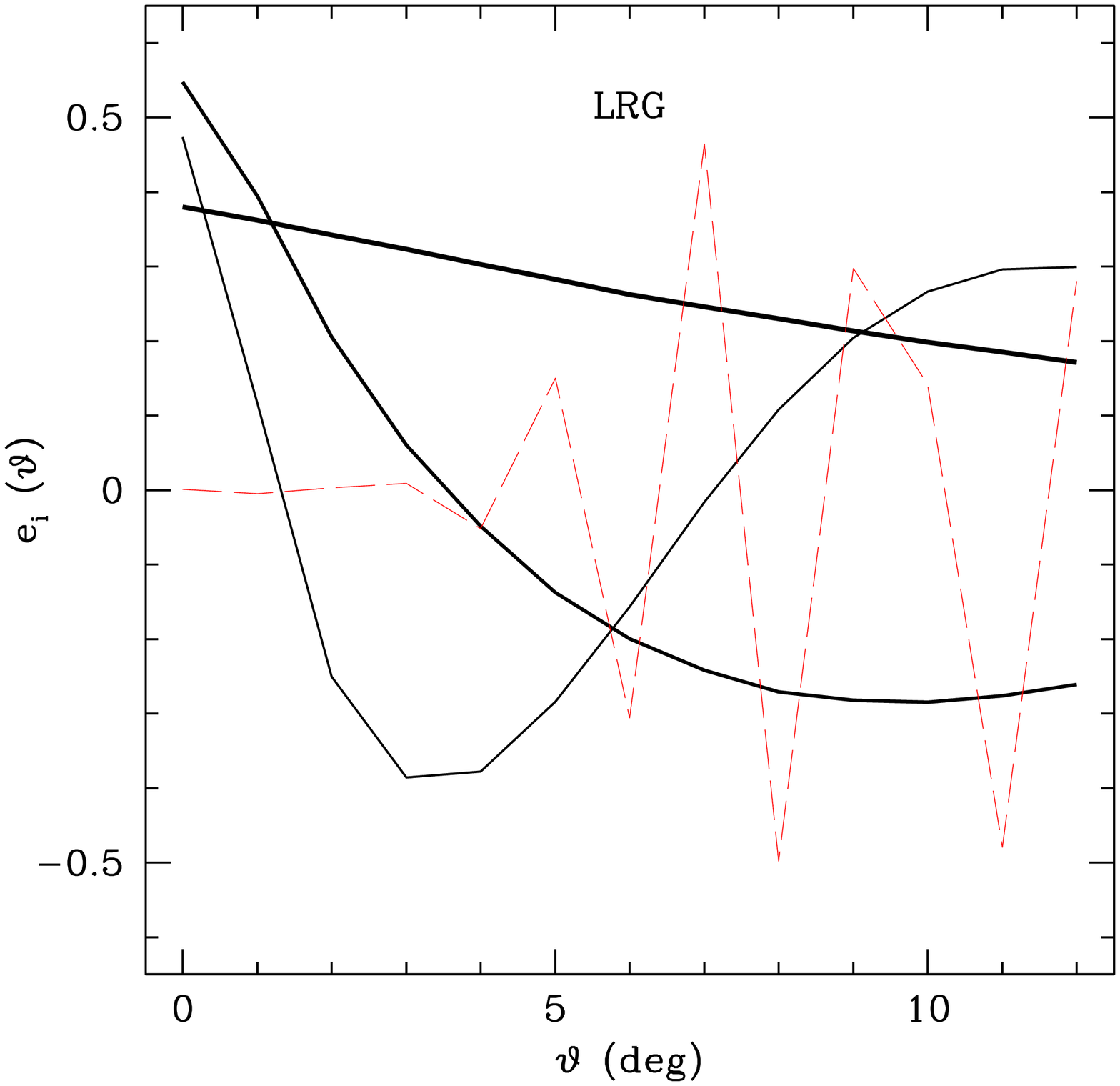}
\caption{Eigenvalues of the MC2 covariance matrices of the cross-correlation between the LRG sample and the CMB (top panel), and first three eigenvectors (bottom panel). The red dashed line shows the highest frequency mode. }
\label {fig:eigenvLRG}
\end{center}
\end{figure}

We can factorise the covariance matrix into the form
\be
{\cal {C}}_{ij} = \sum_{k,l = 1}^n U^T_{ik} \Lambda_{kl} U_{lj},
\ee
where $ \Lambda_{ij} = \lambda_i \delta_{ij} $ is a diagonal matrix whose 
elements are the eigenvalues of ${\cal {C}}_{ij}$; the rows of $ U_{ij} $ are
the 13 eigenvectors $ \hat {\bf e}_{i}$ of the covariance matrix.    We plot 
the variances, $\lambda_i  = \sigma_i^2 $, in the top panel of Fig.~\ref {fig:eigenvLRG}, 
and some of the eigenvectors are shown in the bottom panel.  
There, we can see that the modes associated with the biggest variance are the 
low frequency ones, while the low variance modes oscillate significantly.  
This reflects the fact that the greatest differences between the Monte Carlo realisations is in the 
low frequency behaviour of the cross-correlation functions. 

\begin{figure}[ht] 
\begin{center}
\includegraphics[angle=0,width=1.0\linewidth]{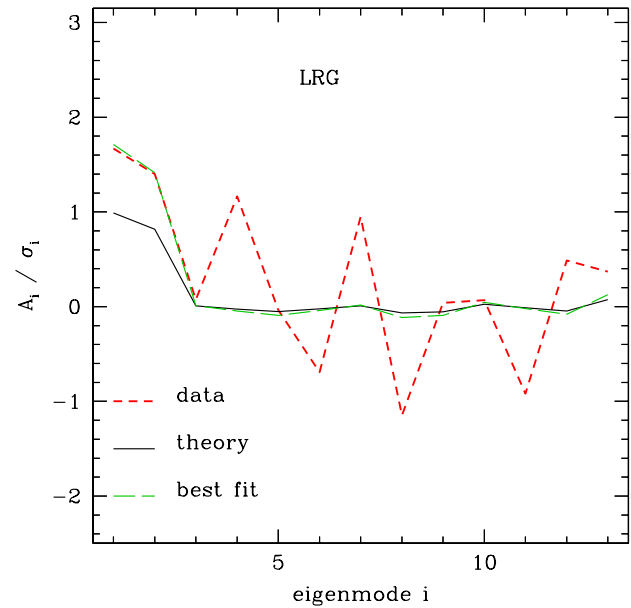}
\caption{Eigenmode decomposition of the amplitude of the measured (red dashed), theoretical (black solid) and best fit (green long dashed) CCF.}
\label {fig:asigmaLRG}
\end{center}
\end{figure}

Both the measured and theoretical CCFs can be
decomposed into this eigenvector basis.
In particular, any cross-correlation vector can be written as 
${\bf v} = \sum_i  A_i  \hat {\bf e}_{i}$, where $A_i \equiv {\bf v}\cdot  \hat {\bf e}_{i}$.  We show in 
Fig.~\ref {fig:asigmaLRG} the decomposition of the data and theory divided 
by the square root of the variance, $ \sigma_i $.   For a typical CCF from the Monte Carlos, these amplitudes 
should be Gaussian distributed with unit variance. 
We can see how the smooth shape of the theoretical real space CCF is 
reflected in this eigenmode decomposition: the theoretical amplitude is very 
well approximated by the first two modes only. However, this is not the case 
for the measured CCF, for which higher frequency modes are also significant. 

\begin{figure}[ht] 
\begin{center}
\includegraphics[angle=0,width=1.0\linewidth]{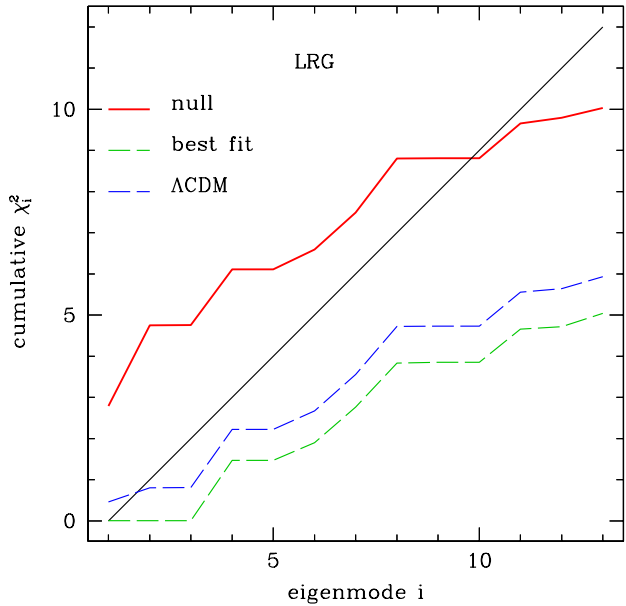}
\caption {Cumulative $ \chi^2_i $ obtained summing the contribution up to the $i$-th eigenmode, for the three models: null hypothesis (red solid), best fit and \LCDM.}
\label {fig:chi2LRG}
\end{center}
\end{figure}

We next look at the contributions to the $ \chi^2 $ from the different eigenmodes.  We show 
in Fig.~\ref {fig:chi2LRG} the evolution of the cumulative $ \chi^2_i $, i.e. 
the cumulative contribution to the $ \chi^2 $ from each eigenmode. 
Here we compare the raw  $\chi^2 $ from the observed cross-correlation function to that for the residuals when 
the theoretical models (\LCDM and the best fit amplitude) are subtracted off.   
As expected, the theoretical models only impact the lowest two eigenmodes.   The low $ \chi^2 $, however, is 
largely the result of the higher frequency modes, which seem to have slightly lower amplitudes than is seen in the Monte Carlos.

If we consider only those two modes which are expected theoretically, 
the $\chi^2 $ for the null hypothesis is actually fairly high: $\chi^2_2 = 4.8$.  
This would exclude the null hypothesis at more than the 90\% level.

\section {Cosmological Constraints} \label {sec:constraints}

Assuming the observed cross-correlations are produced by the ISW effect, we can
compare them with the theory predictions to obtain cosmological constraints.
As described above, the ISW temperature anisotropies are produced as a result of 
time variation in the gravitational potential, and it is the evolution of the potential 
which our measurements constrain most directly.   The cosmological parameters 
which impact the linear evolution of the potential are the dark energy density and its evolution, 
and the curvature of the Universe.  

The actual cross-correlation measurements will also depend on the nature of the large-scale 
structure probe, its spectrum and its bias.  For example, if we normalise to the large scale CMB, 
changing the shape of the power spectrum (e.g., by changing the Hubble constant or the 
dark matter density) will change the variance of the dark matter distribution on smaller scales,  
quantified by $\sigma_8$.   Since the ACFs of the surveys are fixed by observations, changing $\sigma_8$ 
effectively means a different bias will be inferred for each survey.   

The cross-correlations will rise and fall with the amount of structure in the probe.   Thus, instead one could focus on the 
dimensionless cross-correlation, effectively 
\be
r =\frac {C^{T g}_\ell} {\sqrt {C^{TT}_\ell C^{gg}_\ell}} =  \frac {C^{T \delta}_\ell} {\sqrt {C^{TT}_\ell C^{\delta \delta}_\ell}},
\ee
which removes the dependence on bias (assuming it is linear) and probes more directly the ISW effect itself.  
The ISW effect arises on fairly large scales, e.g. $k \sim 0.01 h \rm{Mpc}^{-1}$, depending on the redshift distribution of the survey.  
Equivalently, for each model, we calculate the bias of each survey based on the observed ACF, and use this to find the 
predicted CCF for the model.  

This makes our measurements largely independent of parameters other than $\Omega_{DE}, w, c_{s}$ and $\Omega_k$.
In practise, we choose to keep the dark matter physical density fixed 
$ \omega_m \equiv \Omega_m h^2 = 0.128 $ to the \WMAP best fit value, but the constraints are largely independent of this 
assumption.

\subsection {Models without dark energy}

While many independent probes seem to indicate the existence of dark energy, 
it is worth exploring models which might account for the observations without dark energy; 
recently, an attempt has been made that does this, but which requires a significantly lower Hubble constant, 
modifications to the primordial power spectrum and other non-standard features \cite{Blanchard:2003du}. 
Such models would be dark matter dominated today, and have no late-time ISW effect. 
Our observations of the ISW cross-correlations rule out such models at the $ \sim 4.5 \sigma $ level, based on the difference in the $ \chi^2 $ between the null hypothesis and the \LCDM model in Table \ref {tab:allchi2}.
   Such models also struggle to fit the recent 
observations of the angular scale of baryon oscillations~\cite{Blanchard:2005ev}.

\subsection {Flat \LCDM models}

\begin{figure}[ht] 
\begin{center}
\includegraphics[angle=0,width=1.0\linewidth]{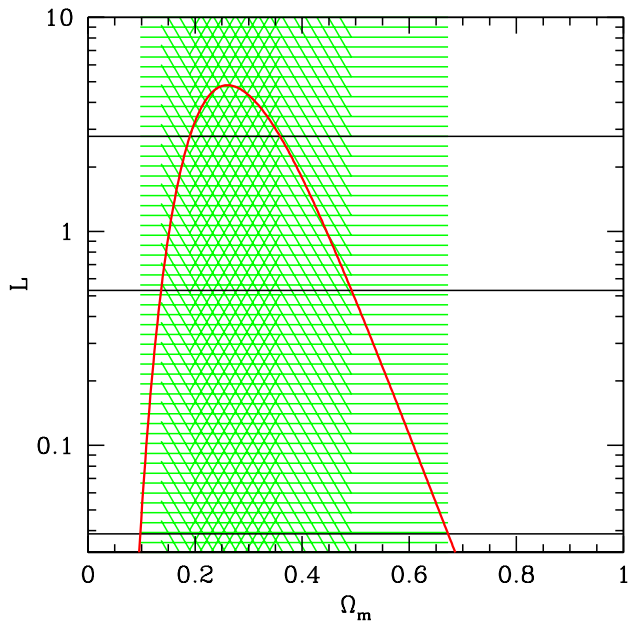}
\includegraphics[angle=0,width=1.0\linewidth]{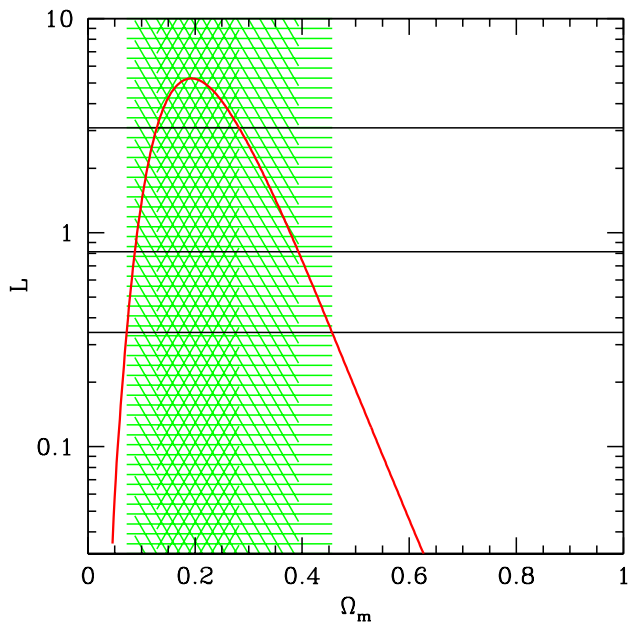}
\caption{Likelihood for flat models with varying $\Omega_{m}$ from the MC1 and MC2 errors. The shaded areas represent 1, 2 and 3 $\sigma$ intervals for $\Omega_{m}$. \LCDM is a good fit to the data.}
\label {fig:lik-Ol}
\end{center}
\end{figure}

Next, we study the likelihood of a family of flat models with varying 
$\Omega_{m}$, $ \Omega_{\Lambda} = 1 - \Omega_{m} $. 
As we can see in Fig.~\ref {fig:lik-Ol},  the \LCDM model is an excellent fit to our data: the 1$\sigma$ 
interval for the parameter is $ \Omega_{m} = 0.26^{+0.09}_{-0.07} $ using the MC1 covariance estimate. 
A higher ISW signal and slightly lower estimate for $\Omega_m$ results from the MC2 errors ($ \Omega_{m} = 0.20^{+0.09}_{-0.07} $); this is due to the higher best fit amplitude in this case.
The error bars can be seen to be very asymmetric, as the ISW effect increases dramatically when the matter density becomes small.  
Models with $ \Omega_{m} < 0.1 $ would predict a much greater cross-correlation than is observed. 

\subsection {Flat wCDM models} 

\begin{figure}[ht] 
\begin{center}
\includegraphics[angle=0,width=1.0\linewidth]{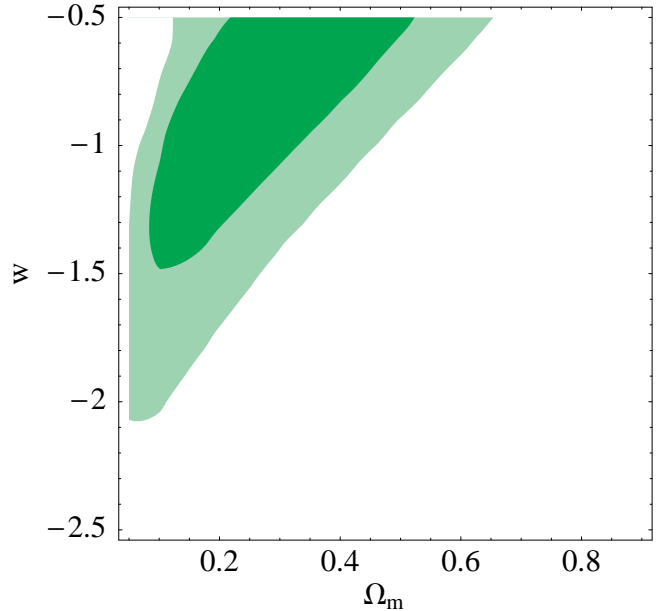}
\includegraphics[angle=0,width=1.0\linewidth]{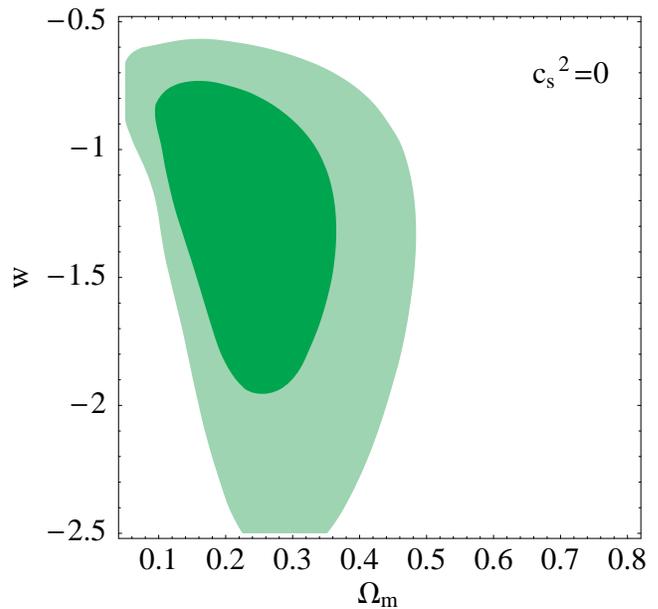}
\caption{Likelihood for flat models with varying $\Omega_{m}$ and $w$ from the MC2 errors. The shaded areas represent 1 and 2 $\sigma$ intervals.  The top panel assumes relativistic sound speed, such as would occur in a quintessence model, while the lower panel assumes 
the opposite extreme of zero sound speed.}
\label {fig:lik-OW}
\end{center}
\end{figure}

We next study the likelihood of a family of flat dark energy models, where we allow the 
dark energy density to evolve with equation of state $w$.  The results are shown in Fig.~\ref{fig:lik-OW}, from which we can see that \LCDM ($w = -1$) is 
very consistent with the measures. We can understand this if we observe that the measured excess in the ISW signal is largely redshift independent, while models along the same degeneracy line with a lower (higher) $w$ would predict an excess at low (high) redshifts respectively.

Initially we assume the dark energy sound speed is $ c_s^2 = 1 $, as is typical in scalar field models like quintessence. 
We also show the same range of models, but with a different dark energy 
sound speed $ c_s^2 = 0 $ in Fig.~\ref{fig:lik-OW}. We can see that in 
this case the degeneracy line changes direction due to the clustering of dark 
energy. \LCDM is still a good fit to the data, as the cosmological constant likelihoods are not affected by the sound speed, 
and there is no clustering in that case.  

\begin{figure}[ht] 
\begin{center}
\includegraphics[angle=0,width=1.0\linewidth]{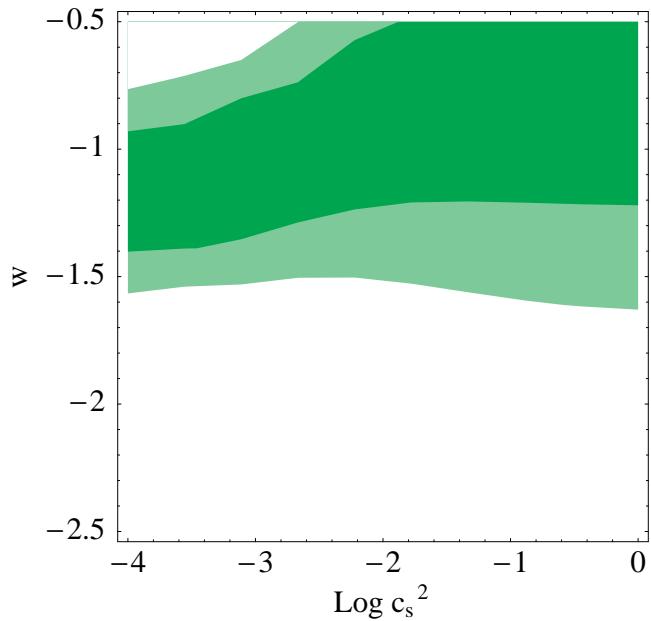}
\caption{Likelihood for flat models with dynamical dark energy as a function of the sound speed, where we fix the matter density based 
on the equation of state, assuming the CMB shift constraint. 1 and 2 $\sigma$ intervals are shown. No constraint is possible for the cosmological constant limit ($w = -1$).}
\label {fig:lik-cs}
\end{center}
\end{figure}

The constraint on the sound speed itself is very weak.  There are too many dark energy parameters (density, equation of state, 
sound speed) to expect any strong constraint.  We reduce the numbers by assuming the CMB shift parameter is fixed to the observed value, coupling 
the equation of state to the dark energy density.   The results can be seen in Fig.~\ref{fig:lik-cs}.   Even with this additional constraint, 
the sound speed is weakly constrained because the data are consistent with a \LCDM model, where there is no dependence on the 
sound speed possible.  

\subsection {Curved \LCDM models}

\begin{figure}[ht] 
\begin{center}
\includegraphics[angle=0,width=1.0\linewidth]{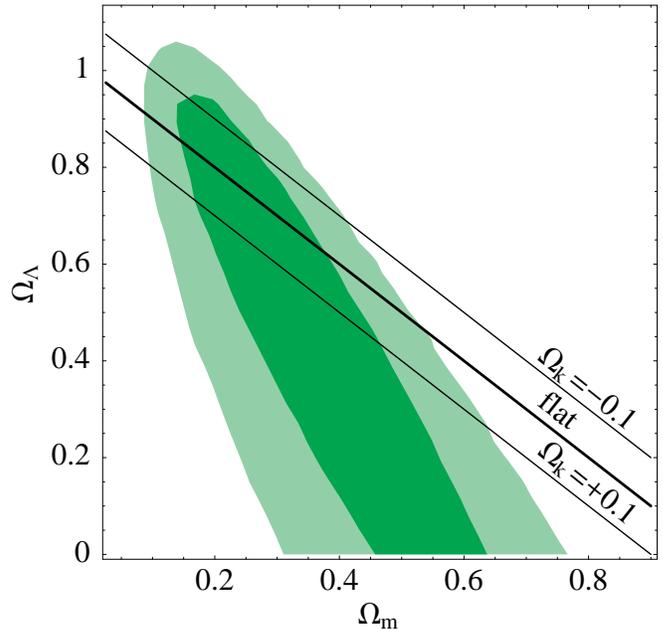}
\caption{Likelihood for curved models with varying $\Omega_{m}$ and $\Omega_{\Lambda}$ from the MC2 errors. The shaded areas represent 1 and 2 $\sigma$ intervals. \LCDM is a good fit to the data.}
\label {fig:lik-OmOk}
\end{center}
\end{figure}

Since curvature can also cause the gravitational potential to evolve, we 
explore the constraints if we relax the flatness condition. However, for simplicity we assume 
the dark energy is a cosmological constant.  We study the likelihood of $ \Omega_m $, with a corresponding 
curvature $ \Omega_k = 1 - \Omega_{\Lambda} - \Omega_m $.   We explore the full 
$ \Omega_m - \Omega_{\Lambda} $ space; we see the relative likelihoods in 
Fig.~\ref {fig:lik-OmOk}, which is obtained with MC2 errors. 

From this figure, we see that \LCDM is still a 
good fit to the data. An interesting feature of this figure is the degeneracy line 
between $ \Omega_m $ and $ \Omega_{\Lambda} $: this is related to the relative 
efficiency of the curvature and dark energy as sources of ISW.
Closed models (above the flat line) give negative ISW, and can cancel the effect 
of increasing the cosmological constant, while the opposite happens for open models (below the flat line).

\subsection {Comparison with other constraints}

Finally, we wish  to compare the ISW constraints to those arising from other cosmological observations, including 
the CMB power spectrum, baryon oscillations and type Ia supernovae.   For the latter, we use measurements of the 
luminosity distance from the Supernova Legacy Survey \cite{Astier:2005qq}. 

\begin{figure}[t] 
\begin{center}
\includegraphics[angle=0,width=1.0\linewidth]{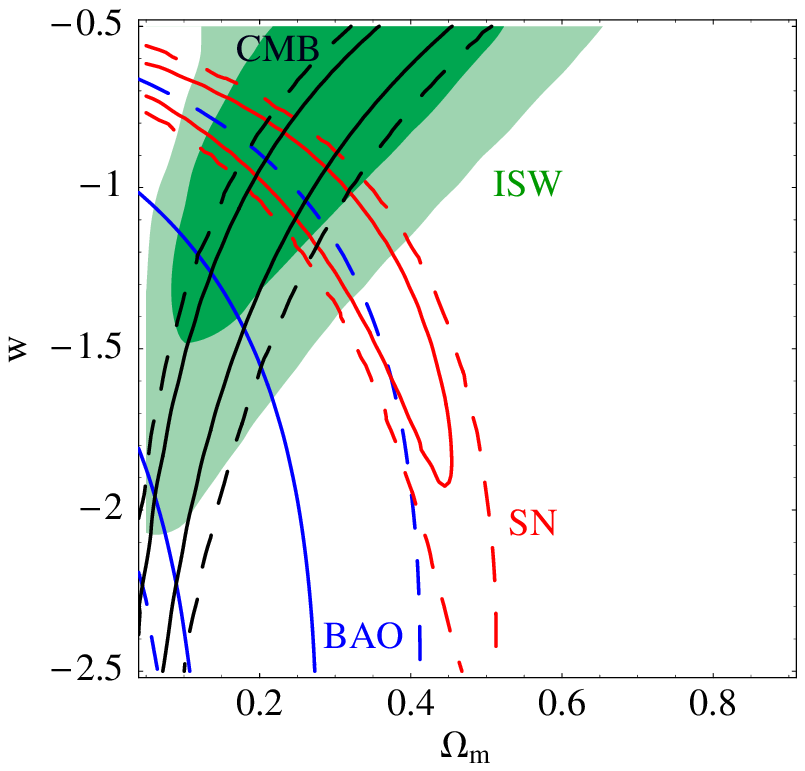} 
\includegraphics[angle=0,width=1.0\linewidth]{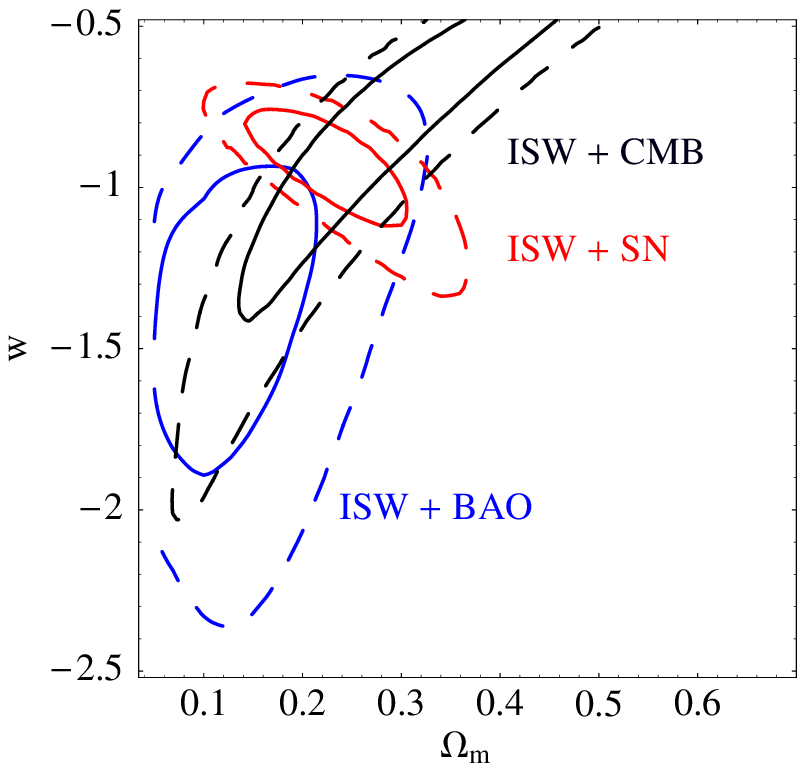} 
\caption{Comparison with constraints from other observations, including CMB shift (black), SNe (red) and BAO (blue) (top panel), and combined likelihoods using the ISW + each one of these other constraints (bottom panel, same colour coding). 1 and 2 $ \sigma $ contours are shown (solid and dashed lines respectively). The MC2 errors are used.}
\label {fig:lik-all}
\end{center}
\end{figure}

For the CMB observations, most of the dark energy information (at least that independent of the ISW effect) is distilled in 
the CMB \emph{shift} parameter, defined as
\be
R \equiv \sqrt {\Omega_m} H_0 \cdot (1 + z_{\star}) \;d_A (z_{\star}),
\ee
where $ d_A(z) $ is the angular diameter distance and $z_{\star}$ is the 
redshift of the last scattering surface ($z_{\star} = 1090$); this expression 
in the flat case reduces to
\be
R = \sqrt {\Omega_m} H_0 \int_0^{z_{\star}}  \frac {dz'} {H(z')} .
\ee
$R$ has been measured to be 
$ R = 1.70 \pm 0.03 $~\cite{Wang:2006ts}. We can see from Fig.~\ref {fig:lik-all} that this 
constraint has a degeneracy direction parallel to the ISW degeneracy in the flat case, 
but is less so in the general curved case.

Finally, for the baryon oscillation (BAO) measurements 
\cite{Percival:2007yw} we use the constraint on the volume 
distance measure defined as
\be
d_V (z) \equiv [(1+z)^2 d_A^2(z) z \; c / H(z)]^{1/3},
\ee
The constraint on this parameter by~\cite{Percival:2007yw} is 
$ d_V (0.35) / d_V (0.2) = 1.812 \pm 0.060 $. 

The SN data is orthogonal to the ISW constraints, and jointly they are 
consistent with the \LCDM model; there is little evidence for additional curvature or 
evolving dark energy.  The CMB shift constraint is similarly consistent
with the cosmological constant concordance model, though the constraints are not as orthogonal to the 
ISW constraints.  The \LCDM model preferred by the SN and ISW measurements is consistent with the CMB shift combined with the 
measurements of the Hubble constant from the HST Key Project  \cite{Freedman:2000cf}.

The exception to this concordance picture comes when the BAO data is considered.  The BAO 
contours are similar to those from the SN, but shifted.  In the flat dark energy case, the combination 
with the ISW prefers a larger dark energy density which has increased with time (phantom).
 When all observations are combined, the BAO data are swamped by the SN data, and the result is 
fully consistent with the concordance model as found by \cite{Percival:2007yw}.

\begin{figure}[t] 
\begin{center}
\includegraphics[angle=0,width=1.0\linewidth]{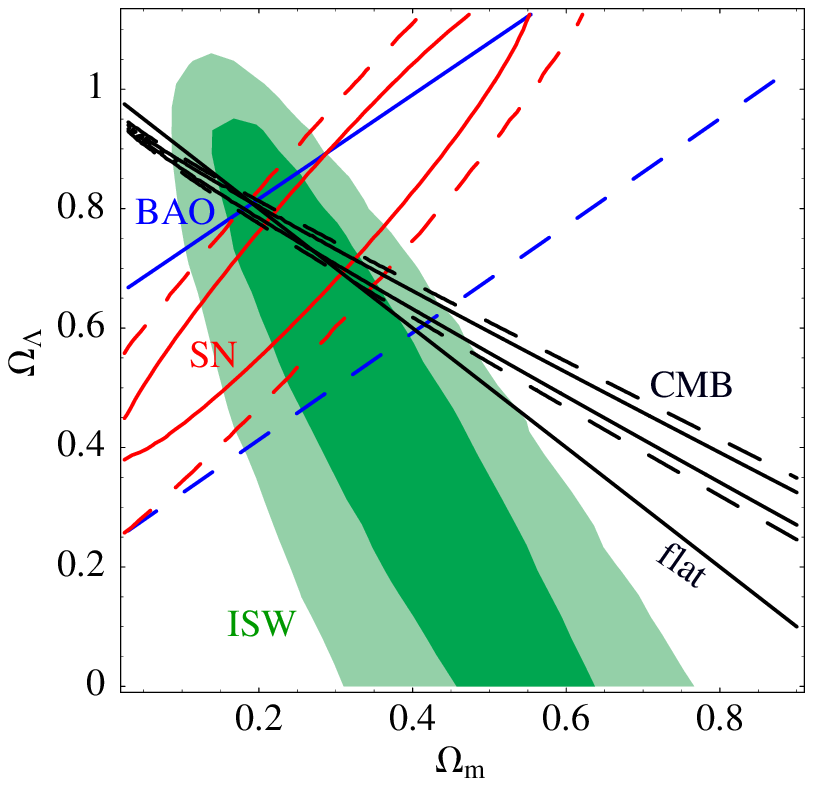}
\includegraphics[angle=0,width=1.0\linewidth]{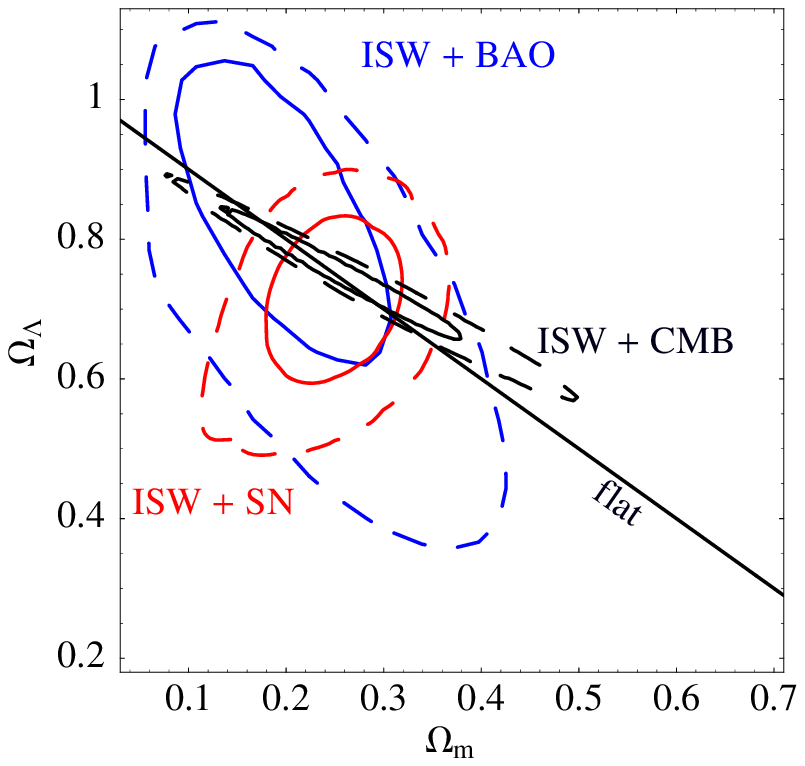}
\caption{Same as Fig.~\ref {fig:lik-all} for the curved case.}
\label {fig:curvlik-all}
\end{center}
\end{figure}

\section {Conclusions} \label {sec:conclusion}

In this paper, we have measured the cross-correlation between the CMB and a large range of 
probes of the density in a consistent way, and have calculated their covariance taking into account 
their overlapping sky coverage and redshift distributions.  While individual measurements vary somewhat 
depending on how the data are cleaned and how the covariance is calculated, the overall significance of 
the detection of cross-correlations is at the $\sim 4.5 \sigma$ level.

These observations provide important independent evidence for the existence and nature of the dark energy.
The observed cross-correlations are consistent with the expected signal arising from the integrated Sachs-Wolfe effect in the
concordance model with a cosmological constant.   The observed signal is slightly higher than expected, higher than the expectation from WMAP best fit model by about $1 \sigma$, thus favouring models with a lower $ \Omega_m$. However, we do not see any significant trend for the excess as a function of redshift, and so there is no indication of an evolving dark energy density.
By combining these results with other cosmological data, we find a generally consistent picture of the behaviour of the Universe, which is converging towards the \LCDM model although the  uncertainties remain considerable. The only partial exception to this picture is the BAO result which, even when combined with our ISW measurement, is in slight tension with the \LCDM model (at $ \sim 1 \sigma$).

The results of our analysis and the covariance matrices are available upon request from the authors (contact \texttt{tommaso.giannantonio@port.ac.uk}).

\section*{Acknowledgements}

We thank Elisabetta Majerotto, Will Percival, Bj{\"o}rn Malte Sch{\"a}fer and 
Jussi V{\"a}liviita for useful conversations. We also thank Filipe Abdalla, Chris Blake and Ofer Lahav for the 
use of the DR6 MegaZ data, and Anna Cabr\'e and Enrique Gazta\~naga for useful discussions 
relating to their earlier measurements. 
We also acknowledge Pier-Stefano Corasaniti and Alessandro Melchiorri who collaborated on the initial version of the modified Boltzmann code, that is publicly available at \texttt {http://www.astro.columbia.edu/\~{}pierste/ISWcode. html}. 
We acknowledge support by a grant from STFC.

While this work was in the final stages of preparation, a similar work appeared by Ho et al. \cite{Ho:2008bz} which 
re-examines some of the same data sets using a multipole space approach.

\clearpage

\appendix

\section {Correlated Monte Carlo maps} \label {app:maps}

\subsection{Basics} 
Here we describe how to make Gaussian maps with a prescribed set of auto- and cross-correlation functions for use in 
the estimation of covariance matrices.   
Let us assume we have $n$ maps, which could include temperature and various density maps at different redshifts or frequencies.  Let us call these maps ${\bf m}_i$ where $i$ ranges from $1$ to $n$. 

Any two maps, ${\bf m}_i$ and ${\bf m}_j$, will be correlated and these correlations will be described by a correlation function $C^{ij}(\vartheta)$ and associated multipole moments $C^{ij}_{\ell}$.  These correlations will be symmetric under interchange of the maps, $C^{ij}(\vartheta) =  C^{ji}(\vartheta)$, so we have $n(n+1)/2$ correlation functions or spectra which describe the two maps.  

Most map making algorithms, like {\tt synfast}\cite{Gorski:2004by}, work in Fourier or spherical harmonic space.  Effectively every mode is given a random amplitude $\xi$, which is a complex number with unit variance and zero mean: $\langle \xi \xi^*\rangle = 1$ and  $\langle \xi \rangle = 0.$   These are then multiplied by the square root of the power spectrum in order to ensure the proper correlation functions.  (There are additional constraints to preserve the reality of the fields on the lattice, e.g. 
$\xi_{\bf k} = \xi_{-\bf k}^*$, but it is not necessary to go through these here). 

It is sufficient to consider a single mode or harmonic amplitude of each map, as all the others will be similar but independent.  Assuming we are working with spherical harmonics, we want to ensure that 
\begin{equation}
\langle a^i_{\ell m}  a^{j*}_{\ell' m'} \rangle = C^{ij}_\ell \delta_{\ell \ell'} \delta_{mm'}.   
\end{equation}
The $\delta$ functions follow simply from using uncorrelated random amplitudes for each harmonic mode.  
For a single map, the right power spectrum is ensured by simply using 
\begin{equation}
a^i_{\ell m} = \sqrt{C^{ii}_\ell} \xi 
\end{equation}
and this is effectively the prescription used by {\tt synfast}.  

When considering more maps, it is necessary to use more random phases, building the final maps from a combination of different maps.  With $n$ maps, $n$ different phases are required for each mode.  Here, we denote the different phases with Latin letters, $a, b, c, ...$  Different phases will be assumed uncorrelated, so that $\langle \xi_a \xi_{a'}^*\rangle = \delta_{aa'}.$

The simplest example is to consider two correlated maps, ${\bf m}_1$ and ${\bf m}_2$.  These are described by three spectra: $C^{11}_\ell, C^{12}_\ell$ and $C^{22}_\ell$.  These are made using the amplitudes 
\begin{eqnarray}
a^1_{\ell m} & = & \xi_a \sqrt{C^{11}_\ell}  \nonumber \\
a^2_{\ell m} & = & \xi_a C^{12}_\ell/\sqrt{C^{11}_\ell}  + \xi_b \sqrt{C^{22}_\ell - (C^{12}_\ell)^2/C^{11}_\ell }. 
\end{eqnarray}
It is simple to verify that with these amplitudes, $\langle a^1_{\ell m}  a^{1*}_{\ell m} \rangle = C^{11}_\ell$, 
$\langle a^1_{\ell m}  a^{2*}_{\ell m} \rangle = C^{12}_\ell$ and $\langle a^2_{\ell m}  a^{2*}_{\ell m} \rangle = C^{22}_\ell$.  

This is simple to implement with \texttt {synfast}.  First create a map with power spectrum $C^{11}_\ell$, and then make a second map using the \emph{ same seeds} and power spectrum  $(C^{12}_\ell)^2/C^{11}_\ell$.  Add this second map to a third map made with a new seed and with power $C^{22}_\ell - (C^{12}_\ell)^2/C^{11}_\ell$.   Note that this should never require taking the square root of a negative number; however, if its very strongly correlated, numerical errors could cause problems.  However, for the weak correlations considered here, this is never an issue.  

The only difficulty is that this inherently produces positive correlations, as the default of the {\tt synfast} code. 
This can be worked around simply.  For example, if $C^{12}_\ell$ is always negative, one can simply flip the signs of the second map after it is produced.  If instead $C^{12}_\ell$ changes sign, then break up the power spectrum into positive and negative pieces, making a map for each and subtracting the `negative' map from the 'positive' map.

\subsection{The general case} 

Next we consider an arbitrary number of maps. 
For simplicity, we drop the $\ell$ and $m$ subscripts where the meaning is unambiguous.  
Effectively, the challenge is to solve for a particular set of amplitudes ${\bf T}$, where 
\begin{eqnarray}
a^1 & = & \xi_a T_{1a} \nonumber \\
a^2 & = & \xi_a T_{2a} + \xi_b T_{2b}  \nonumber \\
a^3 & = & \xi_a T_{3a} + \xi_b T_{3b} +  \xi_c T_{3c} \nonumber \\
a^4 & = & \xi_a T_{4a} + \xi_b T_{4b} +  \xi_c T_{4c} + \xi_d T_{4d} 
\end{eqnarray}
etc., subject to the constraints that $\langle a^i  a^{j*} \rangle = C^{ij}.$

One thus has $n(n+1)/2$ equations with the same number of unknowns $T$.   These begin as: 
 \begin{eqnarray}
C^{11} & = & T_{1a}^2 \nonumber \\
C^{12} & = & T_{1a} T_{2a} \nonumber \\
C^{22} & = & T_{2a}^2 + T_{2b}^2 \nonumber \\
C^{13} & = & T_{1a} T_{3a} \nonumber \\
C^{23} & = & T_{2a} T_{3a} + T_{2b} T_{3b} \nonumber \\
C^{33} & = & T_{3a}^2 + T_{3b}^2 +  T_{3c}^2 \nonumber \\
C^{14} & = & T_{1a} T_{4a}  \nonumber \\
C^{24} & = & T_{2a} T_{4a} + T_{2b} T_{4b} \nonumber \\
C^{34} & = & T_{3a} T_{4a} + T_{3b} T_{4b} + T_{3c} T_{4c} \nonumber \\
C^{44} & = & T_{4a}^2 + T_{4b}^2 +  T_{4c}^2 + T_{4d}^2 \nonumber \\
\end{eqnarray}  
etc.  While quadratic, these can be solved in stages linearly.  Solve the first for $T_{1a} = \sqrt{C^{11}}$.  Use the second to show, $T_{2a} = C^{12}/\sqrt{C^{11}}$ and the third to get $T_{2b} = \sqrt{C^{22} - (C^{12})^2/C^{11}}.$ 
This reproduces what was shown above.  

After this, things continue similarly.  At each point, we use the next equation to solve for the next missing variable: 
\begin{eqnarray}
T_{1a} & = & \sqrt{C^{11}} \nonumber \\
T_{2a} & = & C^{12}/\sqrt{C^{11}} \nonumber \\
T_{2b} & = & \sqrt{C^{22} - (C^{12})^2/C^{11}} \nonumber \\
T_{3a} & = & C^{13}/\sqrt{C^{11}} \nonumber \\
T_{3b} & = & (C^{23} -  C^{12}C^{13}/C^{11})/\sqrt{C^{22} - (C^{12})^2/C^{11}} \nonumber \\
T_{3c} & = & \left[{C^{33} - (C^{13})^2/C^{11} - {(C^{23} - C^{12}C^{13}/C^{11})^2 \over C^{22} - (C^{12})^2/C^{11}}}\right]^{1/2} \nonumber \\
T_{4a} & = & C^{14}/\sqrt{C^{11}} \nonumber \\
T_{4b} & = & (C^{24} -  C^{12}C^{14}/C^{11})/\sqrt{C^{22} - (C^{12})^2/C^{11}} 
\end{eqnarray}
etc.  Things will take similar forms as one goes on, but getting progressively more complicated.  

It can also be programmed recursively, which may be simpler to implement.  By this, we mean, 
\begin{eqnarray}
T_{1a} & = & \sqrt{C^{11}} \nonumber \\
T_{2a} & = & C^{12}/T_{1a} \nonumber \\
T_{2b} & = & \sqrt{C^{22} - T_{2a}^2} \nonumber \\
T_{3a} & = & C^{13}/T_{1a} \nonumber \\
T_{3b} & = & (C^{23} - T_{2a} T_{3a})/T_{2b} \nonumber \\
T_{3c} & = & \sqrt{C^{33} - T_{3a}^2 - T_{3b}^2} \nonumber \\
T_{4a} & = & C^{14}/T_{1a} \nonumber \\
T_{4b} & = & (C^{24} - T_{2a} T_{4a})/T_{2b} \nonumber \\
T_{4c} & = & (C^{34} - T_{3a} T_{4a} - T_{3b}T_{4b})/T_{3c} \nonumber \\
T_{4d} & = & \sqrt{C^{44} - T_{4a}^2 - T_{4b}^2 - T_{4c}^2} 
\end{eqnarray}
etc., with each step using only variables already solved.  
The general recursive expression for these spectra is
\bea
T_{ij}  & = &  \sqrt {C^{ji} - \sum_{k = 1}^{j - 1} T_{ik}^2}, \:\:\:\:\:\:\:\:\:\:\:\:\:\:\: \mathrm{if} \:\: i = j \nonumber \\
&\:& \nonumber \\
T_{ij}  & = &  \frac {C^{ji} - \sum_{k = 1}^{j - 1} T_{ik} T_{jk}} {T_{jj}}  , \:\:\:\:\:\:\: \mathrm{if} \:\: i > j.
\eea

These amplitudes are squared for the input spectra for {\tt synfast}, but one must beware negative cross-correlations as discussed above.  A simple modification to a program like {\tt synfast} could enable it to read in amplitudes rather than spectra, and this would be more efficient compared to reversing the sign of the maps after they are created.

\end {document}